\documentclass[english,graybox]{svmult}
\usepackage{mathptmx}
\usepackage{helvet}
\usepackage{courier}
\usepackage[T1]{fontenc}
\usepackage[latin9]{inputenc}
\usepackage{array}
\usepackage{float}
\usepackage{url}
\usepackage{amssymb}
\usepackage{graphicx}

\makeatletter

\providecommand{\tabularnewline}{\\}

\usepackage{mathptmx}
\usepackage{helvet}
\usepackage{courier}
\usepackage{type1cm}
\usepackage{multicol}
\usepackage[bottom]{footmisc}

\makeindex             

\makeatother

\usepackage{babel}
\begin{document}

\title*{Social-aware Opportunistic Routing:\\
The New Trend
}

\author{Waldir Moreira and Paulo Mendes
}

\institute{Waldir Moreira \at SITI, Universidade Lusófona, Campo Grande, 376,
Ed. U, 1749-024, Lisboa, Portugal\\
 \email{waldir.junior@ulusofona.pt} \and Paulo Mendes \at SITI,
Universidade Lusófona, Campo Grande, 376, Ed. U, 1749-024, Lisboa,
Portugal\\
 \email{paulo.mendes@ulusofona.pt}}

\maketitle

\abstract*{\\\\Since users move around based on social relationships
and interests, the resulting movement patterns can represent how nodes
are socially connected (i.e., nodes with strong social ties, nodes
that meet occasionally by sharing the same working environment). This
means that social interactions reflect personal relationships (e.g.,
family, friends, co-workers, passers-by) that may be translated into
statistical contact opportunities within and between social groups
over time. Such contact opportunities may be exploited to ensure good
data dissemination and retrieval, even in the presence of intermittent
connectivity. Thus, in the last years, a new trend based on social
similarity emerged where social relationships, interests, popularity
and among others, are used to improve opportunistic routing. In this
chapter, the reader will learn about the different approaches related
to opportunistic routing focusing on the social-aware approaches and
how such approaches make use of social information derived from opportunistic
contacts to improve data forwarding. Additionally, a brief overview
on the existing taxonomies for opportunistic routing as well as an
updated one are provided along with a set of experiments in scenarios
based on synthetic mobility models and human traces in order to show
the potential of social-aware solutions.}

\abstract{\\\\Since users move around based on social relationships
and interests, the resulting movement patterns can represent how nodes
are socially connected (i.e., nodes with strong social ties, nodes
that meet occasionally by sharing the same working environment). This
means that social interactions reflect personal relationships (e.g.,
family, friends, co-workers, passers-by) that may be translated into
statistical contact opportunities within and between social groups
over time. Such contact opportunities may be exploited to ensure good
data dissemination and retrieval, even in the presence of intermittent
connectivity. Thus, in the last years, a new trend based on social
similarity emerged where social relationships, interests, popularity
and among others, are used to improve opportunistic routing. In this
chapter, the reader will learn about the different approaches related
to opportunistic routing focusing on the social-aware approaches and
how such approaches make use of social information derived from opportunistic
contacts to improve data forwarding. Additionally, a brief overview
on the existing taxonomies for opportunistic routing as well as an
updated one are provided along with a set of experiments in scenarios
based on synthetic mobility models and human traces in order to show
the potential of social-aware solutions.}

\section{Introduction \label{sec:Introduction}}

The increasing capability of portable devices provide users with new
forms of communication. They can quickly form networks by sharing
resources (i.e., processing, storage) to exchange information and
even share connectivity. This is possible through opportunistic contacts
among devices carried by these users that can forward information
on behalf of other nodes to reach a given destination or connectivity
points.

However, opportunistic communication has to cope with link intermittency.
Due to such intermittency - which results from node mobility, power-saving
schemes, physical obstacles, dark areas (i.e., overcrowded with access
points operating in overlapping channels or no infrastructure at all)
- no end-to-end path may exist, which consequently causes frequent
partitions and high queueing delay.

\begin{figure}[htbp]
\begin{centering}
\textsf{\includegraphics[scale=0.4]{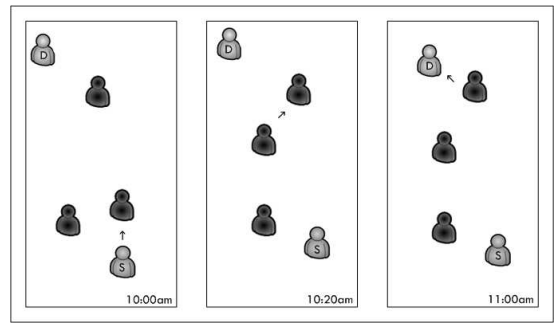}}
\par\end{centering}

\caption{\label{fig:1}Example of opportunistic routing}
\end{figure}

Solutions based on the knowledge of end-to-end paths perform poorly,
and instead, numerous opportunistic routing protocols have been proposed
taking advantage of devices capabilities to overcome intermittency.
Some opportunistic routing protocols use replicas of the same message
to combat the inherent uncertainty of future communication opportunities
between nodes%
\footnote{For the sake of simplicity, node and user are used interchangeably
throughout this chapter.%
}. In order to carefully use the available resources and reach short
delays, many protocols perform forwarding decisions using locally
collected knowledge about node behavior to predict which nodes are
likely to deliver a content or bring it closer to the destination
(cf. Fig. \ref{fig:1}). For that, nodes must have enough processing
power and storage to keep data until another good intermediate carrier
node or the destination is found \cite{rfc4838}, following a store-carry-and-forward
(SCF) paradigm. 

Fig. \ref{fig:1} depicts how content opportunistically reaches its
destination by being transferred and temporarily stored among nodes.

Proposed approaches range from using node mobility to flood the network
for fast delivery (e.g., \emph{Epidemic} \cite{epidemic}) up to controlling
such flooding to achieve the same results based on: encounter history
(e.g., \emph{PROPHET} \cite{prophet,prophet2}), optimized delivery
probability (e.g., \emph{Spray and Wait} \cite{spraywait}), prioritization
(e.g., \emph{MaxProp} \cite{maxprop}), and encounter prediction (e.g.,
\emph{EBR} \cite{ebr}).

Since 2007, a trend has emerged based on different representations
of social similarity: i) labeling users according to their work affiliation
(e.g., \emph{Label} \cite{label}); ii) looking at the importance
(i.e., popularity) of nodes (e.g., \emph{PeopleRank} \cite{people});
iii) combining the notion of community and centrality (e.g., \emph{SimBet}
\cite{simbet} and \emph{Bubble} \emph{Rap} \cite{bubble2011}); iv)
considering interests that users have in common (e.g., \emph{SocialCast}
\cite{socialcast}); v) inferring different level of social interactions
(e.g., \emph{dLife} \cite{dlife}) or predicting future social interactions
(e.g., \emph{CiPRO} \cite{cipro}) from the users\textquoteright{}
dynamic behavior found in their daily life routines. 

These social-aware opportunistic routing solutions have shown great
potential in what concerns information delivery since i) cooperation
among users sharing social aspects (e.g., relationship, tastes, interests,
affiliation, ...) is encouraged, which is beneficial to improve content
dissemination \cite{socialSimilarity}, and ii) social information
is much less volatile than human mobility, providing more robust and
reliable connectivity graphs that aid routing \cite{bubbleTR,bubble2011}. 

Thus, with this chapter the reader is expected to learn about the
different opportunistic routing solutions with emphasis on the social-aware
approaches and how such approaches use social information to improve
data forwarding. Additionally, this chapter provides a brief overview
of existing opportunistic routing taxonomies, so the reader understands
how social similarity has gained attention in the last years. And,
a set of experiments is presented to show that proposals based on
social similarity indeed have a good potential to improve forwarding
in opportunistic networks.

This chapter is structured as follows. Section \ref{sec:Social-oblivious-Opport}
presents the relevant previous work that are oblivious to social information.
Then, the social-aware opportunistic routing approaches are identified
in Section \ref{sec:Social-aware-Opportunistic-Routi} followed by
a brief discussion on the existing taxonomies for opportunistic routing
in Section \ref{sec:Taxonomies}, which makes a reference to when
social similarity started to be considered for routing improvements.
Next, Section \ref{sec:Experiments} presents some results to show
the gains of the social-aware approaches over the social-oblivious
ones, and Section \ref{sec:Conclusions} concludes the chapter.

\section{Social-oblivious Opportunistic Routing Approaches\label{sec:Social-oblivious-Opport}}

According to Jain et al. \cite{oracleknowledge}, deterministic routing
approaches are excellent from a performance point of view, since the
more information a node can get from the network, the wiser its forwarding
decision will be. However, the needed extra knowledge brings more
complexity to the solution and even makes it impossible to be implemented
due to the dynamic nature of user behavior. It was already shown that,
although the optimal solution will need to have a broad knowledge
about the network behavior and traffic demands \cite{oracleknowledge},
even the most simple oracle, called \emph{Contact Oracle}, which contains
information about contacts between any two nodes, is unrealistic as
it is equivalent to knowing the time-varying characteristics of such
networks. Other oracles were defined, including a \emph{Queuing Oracle}
(i.e., information about instantaneous buffer occupancies), and a
\emph{Traffic Demand Oracle} (i.e., information about the present
or future traffic demand), but their assumptions about the network
behavior is even more severe. The existence of such oracles would
support deterministic routing algorithms able to compute an end-to-end
path (possibly time dependent) before messages were actually transmitted.
For example, if a \emph{Contact Oracle} is available, modified Dijkstra
with time-varying cost function based on waiting time would be enough
to find the route. However, the most realistic assumption is that
network topology is not known ahead of time. 

Based on the analysis of deterministic routing approaches, it is clear
that the most suitable solution would be the one based on a local
probabilistic decision, aiming to forward messages based on the opportunities
raised by any contact within range. More elaborated solutions may
also take other information into account to increase the efficiency
of message progression towards a destination. Examples of such extra
information are: history data about encounters, mobility patterns,
priority of information, and social ties. 

In order to understand the importance of social similarity for opportunistic
routing and its applicability, this section provides an overview of
the most relevant proposals to perform opportunistic routing spanning
a 11-year period (2000-2011) and aims at contributing to a broad understanding
of existing opportunistic routing approaches prior to the social-aware
solutions. These routing proposals are grouped into three categories:
single-copy, aiming to improve the usage of network resources; epidemic,
aiming to increase delivery probability; and, probabilistic-based,
aiming to find an optimal balancing between both previous categories.

\subsection{Single-copy Routing}

In resource-constrained networks, which can occur in urban areas with
high spectrum interference, opportunistic routing may lead to waste
of resources when trying to deliver messages to a destination or set
of destinations. Such waste of resources is mainly due to the utilization
of message replication aiming to increase delivery probability. 

Aiming to optimize the usage of network resources, some approaches
avoid replication of messages, forwarding messages to single next-hops
based on available connectivity and some form of mobility prediction.
This means that these proposals perform single-copy forwarding, i.e.,
only one copy of each message traverses the network towards the final
destination. Such copy can be forwarded if the node carrying it decides
(i.e., randomly, or based on a utility function) that another encountered
node presents a higher probability to deliver the message. 

\emph{Minimum Estimated Expected Delay} (MEED) \cite{meed} is an
example of a single-copy forwarding approach that uses contact history
(i.e., connection and disconnection times of contacts) to aid forwarding.
Contact history is a metric that estimates the time a message will
wait until it is forwarded. A per-contact routing scheme is used to
``override'' regular link-state routing decision. That is, instead
of waiting for nodes enclosed in the path with the shortest cost (based
on \emph{MEED} values), it simply uses any other contact opportunity
(i.e., node) that arises prior to what is expected to forward the
message. To be able to do this, \emph{MEED} must recompute routing
tables each time a contact arrives and also broadcast this information
to every other node in the network. 

Spyropoulos et al. (2008) \cite{single-copy-fwd1} present six examples
where this type of forwarding is considered: i) \emph{direct transmission},
where messages are forwarded only to the destination; ii) \emph{randomized
routing}, where messages are only forwarded to encountered nodes that
have forwarding probability $p$, where $0<p\leqslant1$; iii) \emph{utility-based
routing with 1-hop diffusion}, where forwarding takes place if the
utility function - based on encounter timers - of an encountered node
is higher than the one of the current carrier; iv) \emph{utility-based
routing with} transitivity, in which an utility function towards a
destination is updated also considering intermediate nodes that have
high utility to such destination; v) \emph{seek and focus routing},
where \emph{randomized routing} is used to find the best starting
point towards the destination and henceforth a utility-based approach
is used to find the destination; and vi) \emph{oracle-based optimal
algorithm}, with which the future movement is known beforehand allowing
optimal forwarding decisions to be taken aiming to delivery messages
in a short amount of time.

From a resource consumption viewpoint, single-copy forwarding approaches
are quite interesting, since they keep the usage of network (e.g.,
bandwidth) and node (e.g., energy, storage) resources at a low level.
However, they suffer from high delay rates which, consequently, may
result in a low delivery ratio. Another issue is related to the amount
of knowledge that needs to be exchanged/available in order to aid
forwarding, which in some scenarios generates too much overhead and
may be impossible to implement.

To mitigate such problem, the following category of opportunistic
routing approaches relies only on current contacts to increase delivery
probability.

\subsection{Epidemic Routing}

Ubiquitous communication is a feature that has been present in our
everyday life and requires messages to be delivered with high probability
to their destinations, with the help of intermediary nodes able to
implement the SCF paradigm. 

Since networks are created by people moving around, opportunistic
contacts are considered to increase delivery probability in networks
with the aforementioned characteristics in a proposal called \textit{Epidemic
}\textit{\emph{\cite{epidemic}.}}\textit{ }\textit{\emph{With the
}}\textit{Epidemic}\textit{\emph{ routing approach,}} every node in
the network gets at least a copy of each message. Such a full replication
strategy leads to an increase of the delivery rate. Replication of
messages is done by means of summary vectors that are exchanged between
nodes upon a contact. Such summary vectors contain the list of messages
each node is carrying, allowing nodes to exchange all messages that
the other node is lacking. The proposal indeed increases delivery
rate, since every potential forwarder has, with high probability,
a copy of the message, assuming contacts with significant duration
and sufficient buffer space in each node. However, to avoid waste
of resources, each host sets a maximum buffer size that it is willing
to allocate for epidemic message distribution. In general, nodes drop
older messages in favor of newer ones upon reaching their buffer\textquoteright{}s
capacity. This means that the efficiency of the delivery process depends
upon the configured buffer space, which may be substantially improved
with the usage of Bloom filters \cite{bloom}. In order to avoid messages
to be replicated indefinitely, a hop count field determines the maximum
number of epidemic exchanges that a particular message is subject
to, being messages dropped based on the locally available buffer space.
Since the number of hops towards the destination is not known in advance
setting a hop count may decrease the delivery probability. A stale-data
removal mechanism could be more efficient by removing messages that
were already delivered.

In an attempt to avoid waste of network resources, other proposals
emerged based on a controlled replication approach. That is, the number
of nodes which get a message copy is reduced by probabilistically
choosing next nodes or by using a utility function.

\subsection{Probabilistic-based Routing}

Probabilistic approaches are based on the estimation/prediction of
what is the next best set of carriers for each message based on some
probability metric aiming to maximize delivery probability. Probabilistic
forwarding protocols require node mobility patterns that exhibits
long-term regularities such that some nodes consistently meet more
frequently than others over time: the mean inter-meeting time between
two nodes in the past will be close to that in the future with high
probability. 

Within this category, proposals attempt, first of all, to optimize
the delivery probability while avoiding full replication of messages.
Besides this core concern, there are proposals that take into account
the capabilities of nodes, the priority of messages and the availability
of resources aiming to achieve a high delivery ratio with low delay
and resource consumption. Another concern of some probabilistic-based
approaches is to use as few meta-data as possible aiming to decrease
concerns with respect to energy, processing and bandwidth saving.
Having considerable control overhead increases contention in the network
resulting in message discarding and retransmissions.

Since 2003, different delivery probability metrics have been proposed
including frequency encounters \cite{prophet,prophet2,maxprop,prediction,ebr},
aging encounters \cite{fresh,ease,sprayfocus}, aging messages \cite{spraywait,opf},
and resource allocation \cite{prep,rapid}.

\subsubsection{Frequency Encounters}

Proposals based on these metrics have in common the fact that they
rely on the knowledge about how many times nodes meet in a network.
The proposals considered in this sub-category are the \textit{Probabilistic
ROuting Protocol using History of Encounters and Transitivity} (PROPHET),
\textit{MaxProp, Prediction}\textit{\emph{, and }}\textit{Encounter-Based
Routing} (EBR).

One of the most well known approaches, being currently considered
by the \emph{DTN Research Group} (DTNRG) of the \emph{Internet Research
Task Force} (IRTF), is the \textit{Probabilistic ROuting Protocol
using History of Encounters and Transitivity} (PROPHET) \cite{prophet,prophet2}.
\emph{PROPHET} uses a probabilistic metric called delivery predictability,
which indicates what is the likelihood of a node to deliver a message
to a destination based on its past contacts with such destination.
If a pair of nodes does not encounter each other for a while, they
are less likely to be good forwarders of messages to each other, thus
their delivery predictability is reduced in the process. Transitivity
is a property of this predictability where if a node \emph{$A$} regularly
meets a node \emph{$B$} and node \emph{$B$} regularly meets node
$C$, this implies that node \emph{$C$} is also a good node to forward
messages to node \emph{$A$}. Delivery predictability helps to decide
whether the node should replicate or not a given message (It is worth
mentioning that delivery predictability was changed to cope with the
``parking lot'' problem as reported by Grasic et al. (2011) \cite{prophet2}).
Upon a contact, nodes exchange summary vectors that also have information
regarding delivery predictability. This information is used to update
their own delivery predictability vectors, which are used to make
decisions about message forwarding. \emph{PROPHET} delivers messages
only to nodes which are better (in terms of delivery predictability)
than the current carrier node, resulting in a reduction of the consumption
of network resources and a high probability of messages delivery. 

However, flooding can still occur with \emph{PROPHET} if the message
is originated in a node with low delivery predictability (i.e., node
with low mobility) towards the destination, and this node only encounters
nodes with higher delivery predictability.

\textit{MaxProp} \cite{maxprop} is another probabilistic-based approach
that uses a metric called delivery likelihood of messages, by having
each node keeping track of a probability of meeting any other peer.
Using an incremental averaging method, nodes that are seen frequently
obtain higher delivery likelihood values over time. Each time two
nodes meet, they exchange their delivery likelihood probabilities
towards other nodes. Based on the delivery likelihood values computed
by other nodes and by itself, a carrier of a message computes a cost
for each possible path to the destination, up to $n$ hops. The cost
for a path is the sum of the probability of each contact on the path
not occurring. This cost estimation, along with the hop count, are
then used to order messages for scheduling and for dropping. In addition,
\emph{MaxProp} assigns a higher priority to new messages (i.e., low
hop count) to increase their chance of reaching the destination faster,
and tries to prevent reception of the same message twice by including
a hop list in each message, and uses acknowledgments to notify all
nodes about message delivery. Upon contact, two nodes exchange messages
in a specific priority order: first, messages that have these nodes
as final destinations; second, information for estimating delivery
likelihood; third, acknowledgments to remove stale messages; fourth,
messages that have not traversed far in the network; and, fifth, send
messages with highest priority.

By combining the estimation of message delivery likelihood with message
priority and acknowledgments\emph{, MaxProp} is able to reach good
performance regarding message delivery probability and message latency
with transfer opportunities limited in duration and bandwidth. However,
it is considered that nodes have unlimited storage for their own messages
and limited storage for the messages coming from other nodes, which
has an impact on the overall performance.

It is important to mention that the delivery likelihood metric used
in \emph{MaxProp} is different from the delivery predictability metric
employed in \emph{PROPHET} in the sense that the former depends solely
on the probability of nodes to meet each other while \emph{PROPHET}
depends on the probability to meet the destination itself, which means
that \emph{PROPHET} requires more state. However, \emph{MaxProp} requires
nodes to compute possible paths to the destination by concatenation
of delivery probability between nodes while \emph{PROPHET} does not
require further computation, since messages are forwarded if a node
has higher delivery predictability towards the destination than the
carrier.

Song and Kotz proposed a prediction-based approach\textit{ }\textit{\emph{(hereafter
referred to as }}\textit{Prediction}\textit{\emph{) \cite{prediction}
that }}makes use of contact information to estimate the probability
of meeting other nodes in the future. As happens with \emph{PROPHET}
and \emph{MaxProp}, \emph{Prediction} also uses historical contact
information to estimate the probability of meeting other nodes in
the future. However, unlikely previous approaches, \emph{Prediction}
estimates the contact probability within a period of time, based on
a metric called timely-contact probability, which is used to compute
the contact frequency between two nodes, as follows: the contact history
between two nodes $i$ and $j$ is divided into a sequence of $n$
periods of $\Delta T$ starting from the start time ($t_{0}$) of
the first contact in history to the current time. If node $i$ had
any contact with node $j$ during a given period $m$, which is $[t_{0}+m\Delta T,\, t_{0}+(m+1)\Delta T]$,
the contact status of the interval $I_{m}$ is set to 1. The probability
of node $i$ meeting node $j$ in the next $\Delta T$ can be estimated
as the average of the contact status in prior intervals. In \emph{Prediction},
whenever two nodes meet, they exchange the indexes of all their messages.
If the destination of a message is not the node in contact, the probability
to deliver such message through that node is computed. If the probability
of delivering the message via the contacted node (based on the past
average number of encounters with the destination) within a defined
period of time is greater than or equal to a certain threshold, the
message is passed to the node in contact. This proposal presents two
methods for choosing the next node: the decision can be taken by the
node that is sending the message or by the one which is receiving
it. In the former case, a meta-data message is necessary to determine
if the receiver is a good next hop. As for the latter, the receiver
decides whether or not to keep the copy of the message considering
its own probability of coming into contact with the destination.

\emph{Prediction} achieves good performance regarding message delivery
with low number of message transmission and duplication. However,
its performance and storage usage is directly proportional to the
\emph{Time-To-Live} (TTL) allowed for messages which makes it more
suitable for networks tolerant to very long delays (i.e., sparse networks).

The proposal \textit{Encounter-Based Routing} (EBR) \cite{ebr} also
considers the number of times nodes meet in order to predict the rate
levels of future encounters. It simply counts the number of contacts
a node has with other nodes (Current Window Counter) and determines
the Encounter Value (EV) that represents the node\textquoteright{}s
past rate of encounters. The higher EV is, the higher the probability
of successful message delivery. This also determines the number of
replicas of a message that the relay node will get in each contact.
Nodes maintain their past rate of encounters to predict their rate
of future encounters. When nodes meet, they first update their EV
values and estimate the EVs ratio, which is used to determine the
number of tokens of a message replica that will be passed to each
neighbor. This is a kind of time-to-live parameter also used by other
approaches such as \emph{Spray and Focus}. For security reasons, prior
to EV update, nodes exchange information that will aid them to determine
if their EV values are correct.

The prediction of future encounters used by \emph{EBR} allows the
improvement of latency and message delivery by reducing traffic overhead
(i.e., unwanted copies). Messages are only exchanged with nodes that
have high encounter rate, which avoids routes that may not result
in delivery, and minimizes network resource usage. In addition, the
proposal implements a security measure that avoids black hole denial-of-service
attacks from malicious nodes pretending to be part of the network
and announcing fake EV values. However, this approach presents some
drawbacks in scenarios with multiple communities that have low rate
of inter-contact times. In these scenarios, messages may be forwarded
to nodes that indeed have higher EV values, but such messages will
get stuck within the source community. Also, this security measure
incurs in wasting contact opportunities for determining the reliability
of the encountered node, and the proposal may have its performance
degraded in scenarios where nodes have short contact times.

\subsubsection{Aging Encounters}

In this sub-category, the age of encounters are taken into consideration,
and the proposals that fall into it are the \textit{Exponential Age
SEarch} (EASE), \textit{FResher Encounter SearcH} (FRESH), and \textit{Spray
and Focus.}

The \textit{Exponential Age SEarch} (EASE) proposal \cite{ease},
first presented in 2003, was one of the first proposals to consider
history of encounters with a specific destination to support opportunistic
forwarding. In addition to that, \emph{EASE} also considers geographic
position of nodes where a node would make routing decisions based
on the time and location of its last encounter with every other node
in the network. With \emph{EASE,} every node maintains local information
about the time and location of its last encounter with other nodes
in the network. To be considered a good next hop, a node must either
have a more recent encounter with the destination than the current
holder of the message or the node must be physically closer to the
destination. Once the new hop is found, any position-based algorithm
(e.g., DREAM \cite{dream}) can be employed to route the message towards
it. These two phases, namely next hop search and message routing,
are repeated until the best next hop is the destination itself. A
second version of the proposal (\emph{Greedy EASE} \cite{ease}) is
able to change the chosen next hop during the routing phase as it
checks the age of the last encounter with the destination at every
hop.

\emph{EASE} performs quite well in scenarios with nodes presenting
random walk mobility patterns. The performance results show that the
routes towards the destination have the same length (sometimes smaller)
as the optimal case even for very large distances between the communicating
nodes. However, this proposal is highly dependent on the mobility
pattern and destination speed. Its performance is easily degraded
if the destination moves fast turning the solution more costly (i.e.,
in terms of route length). Added to that, if mobile nodes shutdown
their wireless interface (for energy-saving purposes), estimates considering
location are of no use.

The \textit{FResher Encounter SearcH} (FRESH) \cite{fresh} proposal
is an example of a \textquotedblleft{}blind\textquotedblright{} routing
protocol, since it has no notion of coordinates. Each node keeps track
of the time elapsed since the last encounter with every other node,
and uses this information to choose the next hop for message forwarding.
When a sender wishes to initiate data forwarding, it must first search
for the next hop that is determined by the time elapsed since the
last encounter between this potential next hop and the destination.
This search is omni-directional and is done in concentric rings with
increasing radius until the next hop is found. It is necessary that
nodes keep track of their one-hop neighbors to maintain encounter
tables updated.

\emph{FRESH} takes advantage of the time-distance correlation where
the distance traveled during a time interval of duration $t$ is positively
correlated with $t$ (e.g., a node met a few minutes ago is closer
than a node met two hours before). With that, it is able to improve
the performance of route discovery with no need for global knowledge
of the network. Instead, the proposal is based on a distributed implementation
where next hop search is defined in terms of local information (e.g.,
encounter tables). Like with \emph{EASE}, performance depends on nodes
mobility processes as the time-distance correlation becomes noisier
with heterogeneous speeds. That means that a node that has just encountered
the destination may not be close to it if the destination is moving
too fast. Another issue is that \emph{FRESH} may suffer with loops
in routing if source and destination are not part of a connected subset
of nodes. This is indeed a problem especially in scenarios where isolated
cluster of nodes may be formed. There is also an overhead related
to the need of having one-hop neighbor encounter table updated.

\textit{\emph{The}}\textit{ }\textit{\emph{more recent }}\textit{Spray
and Focus} \cite{sprayfocus} approach proposes a scheme were a fixed
number of copies are spread initially exactly as in \emph{Spray and
Wait} \cite{spraywait} (with the subtle difference of using only
$\frac{1}{3}$ or $\frac{1}{2}$ of the $L$ messages normally used
in \emph{Spray and Wait}), but then each copy is routed independently
according to single-copy utility-based scheme with transitivity \cite{single-copy-fwd1}.
In the spraying phase, ideally it would be good to be able to choose
as relays the $L$ nodes that most frequently encounter the destination.
However, waiting for a \textquotedblleft{}better\textquotedblright{}
relay may mean that opportunities to spread extra copies are forfeited.
Hence, the \emph{Spray and Focus} scheme uses a greedy spraying phase
by implementing a binary spraying algorithm to minimize the amount
of time it takes to spray all $L$ copies, moving the problem of looking
for a possibly better relay to the focus phase. In the focus phase,
each potential router maintains a timer for every other node in the
network, recording the time elapsed since the two nodes came within
transmission range for the last time. These timers are similar to
the age of last encounter \cite{fresh} and contain indirect location
information. For a large number of mobility models, it can be shown
that a smaller timer value on average implies a smaller distance from
the node in question. 

\textit{Spray and Focus} outperforms flooding-based (i.e., \emph{Epidemic})
and single-copy schemes (i.e., \emph{Randomized} \emph{Flooding},
and \emph{Utility}-\emph{based} \emph{Flooding}) as well as the other
spraying algorithm (i.e., \emph{Spray and Wait}) under realistic mobility
scenarios (e.g., modeling human behavior), by forwarding messages
to nodes which have a ``closer'' relationship (determined by the
encounter timers) with the destination. Also, \emph{Spray and Focus}
presents good performance in scenarios with heterogeneous mobility
using an algorithm that is able to diffuse timer information much
faster than regular last encounter based schemes. However, since its
performance is highly dependent on the use of encounter timers, it
can be easily degraded in scenarios where nodes are highly mobile
as timers quickly become obsolete.

\subsubsection{Aging Messages}

These proposals have in common the fact that they aim to avoid messages
to be kept being forwarded in the network by creating metrics that
define the age of message copies. \textit{Spray and Wait}\textit{\emph{,
and }}\textit{Optimal Probabilistic Forwarding} (OPF) are the proposals
comprised by this sub-category.

\textit{\emph{One of the first proposals was }}\textit{Spray and Wait
}\textit{\emph{\cite{spraywait}, which}} decouples the number of
transmissions per message from the total number of nodes, generating
a small number of transmissions in a large range of scenarios. Initially,
copies of a message are spread quickly in a manner similar to epidemic
routing. However, \emph{Spray and Wait} stops when enough copies have
been sprayed in order to avoid flooding, while guaranteeing that at
least one copy will reach the destination with high probability. By
exploiting the mobility of nodes, \textit{Spray and Wait }operates
in two steps: first, the source determines a certain number of adjacent
nodes that are going to get a copy of the message (spraying phase).
In the second step, the nodes that got a copy of the message deliver
it directly to the destination when it gets within range (waiting
phase). Each generated message can have \emph{$L$} copies of it distributed
in the network. This number of copies can be determined in two ways:
i) based on the number of nodes \emph{$M$} and size of the network
\emph{$N$}; ii) by estimating \emph{$M$} when both \emph{$M$} and
\emph{$N$} are unknown. Once \emph{$L$} is known, the source can
spread the copies of the message by passing only one copy of the message
(\emph{Source} \emph{Spray and Wait}), or $L/2$ copies (\emph{Binary}
\emph{Spray and Wait}) to each encountered node. In the latter case,
the receiver of $L/2$ copies will spread the obtained copies in the
same way. If the destination is not found in the spraying phase, the
nodes holding one copy of the message will forward it directly to
the final destination. 

\emph{Spray and Wait }exploits node mobility being able to limit the
total number of copies and transmissions per message resulting in
an energy-efficient solution with low delivery delay, although the
achieved delay is inversely proportional to the number of copies.
If nodes move quickly enough around the network, \emph{Spray and Wait}
shows that only a small number of copies can create enough diversity
to achieve close-to-optimal delays. However, there is no acknowledgment
mechanism to get rid of copies of already delivered messages and no
mechanism to select the best set of forwarders in the spraying phase.
In what concerns computational effort, determining \emph{$L$} is
not an easy task, since it is necessary to know \emph{$M$} beforehand
and it depends on nodes performing independent random walks. This
can easily result in an inaccurate measure of \emph{$L$,} which degrades
the algorithm's performance. This problem is even worse in large dense
networks with frequent disconnections and nodes following different
mobility patterns.

The \textit{Optimal Probabilistic Forwarding} (OPF) \cite{opf} protocol
replicates a message upon node encounter if, by doing so, such action
increases the overall delivery probability of such message. That is,
if this action maximizes the joint expected delivery probability of
the copies to be placed in system (i.e., in the sender and receiver
nodes of the message). \emph{OPF} aims to maximize the delivery probability
based on a particular knowledge about the network, relying on the
assumptions that node mobility exhibits long-term regularity (enabling
the estimation of mean inter-meeting times) and that each node knows
the mean inter-meeting time of all pairs of nodes in the network.
\emph{OPF} metric reflects not only the direct delivery probability
of a message, such as in \emph{PROPHET}, but also the indirect delivery
probability when the node can forward the message to other intermediate
nodes, as in \emph{MaxProp}. However, unlike \emph{MaxProp}, \emph{OPF}
metric reflects a hop-count-limited forwarding scheme, based on a
function of two important states of a message: remaining hop-count
and residual lifetime. Such utility function may estimate the effect
that message replication may have on the expected delivery rate while
satisfying the constant on the number of forwardings per message (\emph{OPF}
has a performance awareness as happens with \emph{RAPID}, for instance). 

With \emph{OPF} every message has a residual time-to-live ($T_{r}$)
that also denotes a given meeting time slot, and nodes know the mean
inter-meeting time ($I_{i,j}$) between any two nodes $i$ and $j$
in the network. This is used to determine the meeting probability
($M_{i,j}$) among any two nodes and the delivery probability ($P_{i,j,K,T_{r}}$)
between nodes $i$ and $j$ of a message with \emph{$K$} remaining
hop-count and $T_{r}$ residual time to live. The delivery probability
is simply given by $P_{i,d,0,T_{r}}$ if the message cannot use anymore
hops to reach the destination. However, when the message is at \emph{$K$}
hops from the destination, forwarding will take place if the combined
probability of the two new copies of the message at the next time-slot
$T_{r-1}$ (i.e., $1-(1-P_{i,d,K-1,T_{r-1}})\times(1-P_{j,d,K-1,T_{r-1}})$)
is greater or equals to the probability of not forwarding it at all
(i.e., $P_{i,d,K,T_{r-1}}$). So, when node $i$ meets node $j$,
whether $i$ should forward the copy to $j$ depends on whether replacing
the copy in $i$ with two logically new copies (i.e., in $i$ and
$j$) increases the overall delivery probability. \emph{OPF} also
comes in another version where \emph{$K$} is then substituted by
the number of logical tickets (\emph{$L$}, as in \emph{Spray and
Wait}) which are going to be distributed between the two replacing
copies in a message forwarding.

\emph{OPF} is able to achieve good overall delivery rate with a subtle
increase in delay since it only forwards messages to really good relay
nodes. Its performance is better if the relationship between nodes
is greater and the hop count (\emph{$K$}) allowed for each message
is chosen wisely. However, this good performance comes with a cost,
since it is really dependent on the amount of routing information
available. In networks where only local information is available due
to the dynamicity of nodes, \emph{OPF} will have its performance degraded
since it needs the mean inter-meeting time of all nodes in the network
and such information may be difficult to obtain. Added to that, since
the mobility model considered follows an exponential inter-meeting
time, the measurements may not represent human behavior as it is known
that a power law distribution better represents such behavior \cite{bubbleTR}.

\subsubsection{Resource Allocation}

In what concerns approaches that are aware of available resources,
we have the \textit{Resource Allocation Protocol for Intentional DTN}
(RAPID), and \textit{PRioritized EPidemic} (PREP).

The proposal dubbed \textit{Resource Allocation Protocol for Intentional
DTN} (RAPID) \cite{rapid} opportunistically replicates messages based
upon a utility function that estimates the effect that message replication
may have on a predefined performance metric in a network with resource
constraints. When nodes meet, they exchange meta-data about messages
to be delivered and acknowledgements about already delivered messages,
along with messages destined to each other. With the meta-data, nodes
are able to determine the marginal utility (which must have the highest
increase based on the performance metric considered) of replicating
messages between them.\emph{ RAPID }calculates the effect of replication
considering resources constraints by exchanging meta-data through
an in-band control channel that allows it to have a global state of
the network resources (e.g., length of past transfers, expected meeting
times, list of delivered messages, delivery delay estimate for buffered
messages, changed information on messages since last exchange). 

\emph{RAPID} has a small cost in the usage of contact opportunities
due to the utilization of an in-band control channel. Moreover, the
information exchanged in such channel may not always be updated due
to node mobility, delivery delay, and unacknowledged messages. This
cost may be very high in bandwidth-constrained scenarios with short-lived
contact opportunities. Besides that, there are no overall performance
guarantees since heuristics are based on sub-optimal solutions supported
by one metric at a time. In addition, the performance is also related
to the used mobility pattern (e.g., predictable vehicular movements)
and can be degraded in scenarios with unpredictable mobility patterns. 

In what concerns metric-based approaches, the \textit{PRioritized
EPidemic} (PREP) \cite{prep} proposal is also based on message prioritization
and the idea of prediction. \emph{PREP} introduces the average availability
(AA) metric that measures the average fraction of time a link will
be available in the future (i.e., the inter-node cost) and defines
drop and transmit priorities (in which lower values indicate high
priority) for each message. By using a regular discovery algorithm,
each node finds out about its links towards neighboring nodes. According
to available information about the past state of the link (i.e., up/down),
a node can determine the availability of the links for future use.
Then, costs are assigned to links based on their AA values and epidemically
broadcasted in the network. To find the lowest cost path to a destination
(or its whereabouts), the Dijkstra's shortest path algorithm is employed.
Any changes to AA values will trigger link costs updates, which are
again broadcasted through \emph{Link State Advertisements}.

\emph{PREP} is able to generate a gradient of message replication
density that is inversely proportional to the distance towards the
destination. That is, messages' copies are kept as close as possible
to their destinations. \emph{PREP} sets a drop priority where messages
shall be discarded, upon a full buffer, according to a cost determined
by the distance between the message's holder and destination. The
greater the distance, the higher the drop priority. In addition, a
transmission priority is also set to messages considering their expiry
time and the cost previously mentioned. This allows a wiser usage
of resources (e.g., storage and bandwidth) having a good effect on
message delivery. However, in dynamic scenarios, \emph{PREP} has its
performance degraded since its delivery capability is inversely proportional
to the level of disruption happening in the scenario.

\subsubsection{Considerations}

This section introduced the reader with few social-oblivious opportunistic
in order to show their advantages and disadvantages. With this, we
expect the reader to understand the need to incorporate social information
in opportunistic routing has emerged as nodes area carried by human
who happen to have social similarities (e.g., same workplace, shared
interests, ...). It is important to note that the presented solutions
were chosen according to the: i) number of times they have been referenced
(i.e., served as benchmarks); or ii) number of benchmarks they have
used for their evaluation.

We must highlight that some authors may consider some of the aforementioned
solutions as social-aware (for considering the history of encounters,
for instance). However, it is our belief that social-aware solutions
are those which have much more elaborate utility functions and/or
consider features that can be used to identify/classify individuals
or groups of these, i.e., common affiliations, shared interests, social
ties, popularity, centrality, among others which will be further discussed
in Section \ref{sec:Social-aware-Opportunistic-Routi}\@.

\section{Social-aware Opportunistic Routing\label{sec:Social-aware-Opportunistic-Routi}}

Within the previous solutions for opportunistic routing, one of the
processes that may lead to significant consumption of energy, processing
capability and bandwidth is the prediction of mobility patterns, which
is quite common to all probabilistic routing approaches. One alternative
may be to devise probabilistic solutions that exploit not only mobility
of nodes but also their social similarities. The reason is that mobility
patterns change faster (causing the appearance of unwanted traffic
due to out-of-date information) than social relationships between
people within a society. The fact that social relationships are less
volatile than mobility behavior has been proven \cite{bubbleTR,bubble2011}
to be rather useful in forwarding decisions. Additionally, the higher
social similarity, the better content dissemination is \cite{socialSimilarity}.
Hence, since 2007 several approaches have been investigating the exploitation
of social aspects such as relationships, interests, common affiliations,
in order to improve the delivery rate while decreasing the consumption
of network resources. The proposals that are part of this sub-category
are \emph{Label}, \emph{SimBet}, \emph{Bubble Rap}, \emph{SocialCast},
\emph{PeopleRank}, \emph{dLife,} and \emph{CiPRO}.

\textit{\emph{The }}\textit{Label} \cite{label} approach was one
of the first proposals to employ social characteristics into opportunistic
routing. Experiments were conducted in INFOCOM 2006 where nodes were
labeled telling others about their affiliation/group, and this allowed
nodes to forward messages directly to destinations, or to next hops
belonging to the same group (i.e., same label) as the destinations.
The proposal looked not only to inter-contact time distribution for
all the nodes inside a group but also to the inter-contact time distribution
between two groups (i.e., friendship ties) with results showing that
nodes from one group may be good forwarders for nodes in the corresponding
friendship group. The \emph{Label} proposal provided the first indication
that exploiting social similarities improves delivery ratio and especially
delivery cost (i.e., total number of messages and duplicates transmitted).
Another observation is that friendship between different communities
(i.e., unusual connections among nodes of both communities) can be
used to slightly improve delivery ratio.

\emph{Label}'s performance is directly related to the allowed message
TTL and the mixing rate of nodes in the scenario. Delivery ratio is
very low in the case of messages with short TTL, and it is easily
degraded if nodes do not mix well. The reason is that \emph{Label}
performs only one-hop delivery and just to nodes belonging to the
same community as the destination, which means that delivery may fail
if the sender never encounters members of the same community as the
destination.

\textit{\emph{Some network nodes may have such a behavior that make
the usage of encounters inefficient to forward messages, due to their
sporadic meeting rates. For instance, a node may be involved in a
highly clustered network in which none of the nodes have directly
or indirectly met the destination node. However, paths between clusters
may be insured by nodes that form bridges based on weak acquaintance
ties. In this context, the }}\textit{SimBet}\textit{\emph{ \cite{simbet}
approach proposes to forward data based on the identification of these
bridges and the identification of nodes that reside within the same
cluster as the destination node. The major contribution is a new forwarding
metric based on ego network analysis to locally determine}} betweenness
centrality of nodes (i.e., importance of a node in the system, defined
as the number of connections between nodes belonging to different
communities that cross the referred node) and social similarity (i.e.,
probability of future collaboration between nodes in the same community).
These two social parameters are determined from an adjacency matrix
that each node keeps to track contacts (direct and through neighbor
nodes) with other nodes in the network. When a node $i$ meets a node
$j$, they exchange messages they have to each other and request each
other's list of contacts. With this list, they can update their own
contact list along with their betweenness ($Bet$) and similarity
($Sim$) values. Then, they exchange summary vectors that contain
the destinations to which they are carrying messages along with their
updated $Bet$ and $Sim$ values. For each destination in the summary
vector, they determine their \emph{SimBet} utility. With this, a vector
of destinations is created containing all the destinations to which
the node has highest \emph{SimBet} utility. This vector of destinations
is exchanged and nodes exchange messages to destinations present in
their own vector and remove such messages from their buffers.

\emph{SimBet} is able to forward messages even if the destination
node is unknown to the sending node or its contacts. In this case,
the message is routed to a structurally more central node where the
potential of finding a suitable carrier is much higher. Moreover,
\emph{SimBet} makes no assumptions about the control of node movements
or knowledge of node future movements. Finally, with \emph{SimBet}
messages are forwarded solely based on locally obtained information.
It works based on forwarding a single copy of each message in the
network, which makes it able to reduce resource consumption, mainly
buffer space and energy. It has good overall performance regarding
message delivery, which is close to \emph{Epidemic}'s but with highly
improved delivery cost (i.e., very low number of required forwards
to reach destination). However, this proposal may suffer with high
delay since the level of contact (i.e., how often and with whom nodes
meet)\textbf{ }between nodes is a key aspect regarding dissemination
of information. That is, if this level of contact between nodes is
low, information (e.g., \emph{$Sim$} and \emph{$Bet$} values, contact
lists) will take longer to be updated and diffused. Another issue
regarding performance is the contact time, which can have a strong
influence especially in scenarios where contacts are short lived.

Another proposal known as \textit{Bubble Rap} \cite{bubblerap,bubble2011}
also uses node centrality along with the concept of community structure
to perform forwarding. With \emph{Bubble Rap,} nodes are grouped based
on social parameters (i.e., number of contacts and contact duration)
and have a local/global popularity index (obtained from betweenness
centrality). Messages within the same community are forwarded using
local popularity whereas messages traversing different communities
use a combination of local/global popularity to reach the final destination.
In the second case, whenever the message is forwarded to a member
of the destination\textquoteright{}s community, the current carrier
deletes it from the buffer to prevent further dissemination. The algorithm
employed in this proposal is rather simple. If a source node wishes
to send a message, all it needs to do is to check if the community
of the destination node is the same as its own. If so, for every encountered
intermediate node, it compares their local ranks and generates a copy
to the encountered node that has higher rank value. Otherwise, it
passes this message copy to an encountered node belonging to the same
community as the destination or having higher global rank value.

\emph{Bubble Rap} considers that nodes belong to different size communities
and that such nodes have different levels of popularity (i.e., rank).
With this, it can mimic human relationship allowing it to achieve
good overall performance regarding delivery success rate with acceptable
delivery cost. The performance is even better as the number of different
communities and message TTL increase showing its capability to deal
with human social behavior. However, reaching a destination in a different
community is quite exhaustive, especially if the source node has the
lowest rank and its community has a high number of nodes. This will
incur undesirable replication within the source community. Moreover,
centrality can result in overloading (e.g., processing, energy, buffer)
popular nodes since they are outnumbered compared to the number of
global nodes. In addition, it does not always mean that a high centrality
node has the best contact probability with the destination community.

\textit{\emph{Differently from previous proposals, }}\textit{SocialCast}
\cite{socialcast} shows that forwarding can be achieved not only
based on the social ties and mobility patterns, but also considering
the interests of destinations. The proposed routing protocol determines
a utility function based on the predicted node\textquoteright{}s co-location
(i.e., probability of nodes being co-located with others sharing the
same interest) and change in degree of connectivity (i.e., related
to mobility and representing changes in neighbor sets), which is used
to calculate how good data carrier a node can be. This proposal is
based on the publish-subscribe paradigm, that is, nodes publish content
on the network that is received by nodes according to their subscribed
interests. Nodes get copies if they have higher utility regarding
a given interest than the node currently carrying messages with content
matching such interest. The proposal comprises three phases: i) \emph{interest
dissemination}, in which each node broadcasts, to its first-hop neighbors,
its list of interests along with its updated utilities regarding its
interests as well as the last received messages; ii) \emph{carrier
selection}, in which if the utility function of a neighbor node regarding
a given interest is higher than the current carrier node, this neighbor
is selected as the new carrier; and, iii) \emph{message dissemination},
in which messages are replicated to interested nodes and/or passed
to the new carrier.

\emph{SocialCast} allows messages to reach their destinations with
a very low number of replications (i.e., reduced resource consumption)
and stable latency. The result is good delivery ratio with low TTL
values as messages are delivered within few hops. However, the co-location
assumption (i.e., nodes with same interests spend quite some time
together) may not always be true \cite{people}. Since such assumption
is of great importance, the proposal is compromised in scenarios where
it does not always apply.

Also considering node mobility and social interaction, \emph{PeopleRank}
\cite{people} makes use of stable social information between nodes
to decide on forwarding. As its own name suggests, \emph{PeopleRank}
sets ranks to nodes according to their social interaction, and use
this ranking to decide on the next hop for data exchange as it is
known that socially well-connected nodes become the best forwarders
for message delivery. This ranking process is analogous to Google's
page rank system in which the relative importance of a Web page is
determined according to its links to/from a set of pages.

Thus, nodes are ranked according to their position in the social graph,
i.e., considering their linkage to other important nodes in the network.
In order to determine its rank, a node needs to be acquainted to its
socially connected neighbors and their respective ranks. So, when
two nodes meet, they exchange their ranks and neighbor sets, update
their own ranks, and exchange messages according to the new determined
ranking. The node with the highest rank gets the messages.

The more social information is available to nodes, the better is the
overall performance regarding delivery success rate. \emph{PeopleRank}
is also able to keep the cost associated to message delivery very
low and with short delays. However, it is proven that considering
only socially connected nodes is not enough to guarantee a good performance
level, since socially disconnected nodes are also able to forward
messages and could be considered to improve performance.

Most of the previous solutions focus solely inter-contact \cite{impactHuman},
and still significant investigation is required to understand the
nature of such statistics (e.g., power-law, behavior dependent on
node context) \cite{inter-contact}. Another drawback of such approaches
is the instable proximity graphs they create which follows mobility(encounter)-based
social similarity metrics \cite{bubble2011}. Instead of considering
the dynamicity of mobility, the evolving feature of social network
structure of nodes as they move around meeting different nodes throughout
the day is what matters most. And such feature has been shown to be
imperative when building proximity graphs based on social interactions
\cite{thyneighbor}.

Existing solutions \cite{label,simbet,socialcast,bubble2011,people}
succeed in identifying similarities (e.g., affiliation, communities,
interests) among users, but their performance is affected as dynamism
derived from users' daily routines is not considered.

With \emph{dLife} \cite{dlife}, the dynamism of users' behavior found
in their daily life routines is considered to aid routing. The goal
is to keep track of the different levels of social interactions (in
terms of contact duration) nodes have throughout their daily activities
in order to infer how well socially connected they are in different
periods of the day. 

The assumption here is that the time nodes spend together can be used
as a measure of the strength of the social ties among them. To achieve
that, \emph{dLife} defines two utility functions: Time-Evolving Contact
Duration (TECD) that measures the level of social interaction (i.e.,
social strength, $w(a,b)$) among pairs of nodes during their daily
routine activities; and TECD Importance (TECDi) that measures the
user\textquoteright{}s importance, $I(x)$, considering its node degree
and the social strength towards its neighbors. Upon a contact, nodes
exchange their social weights towards other nodes as well as their
importance, and replication only occurs if the encountered node has
a stronger social with the message's destination or higher importance
in the system than the current message's carrier for that specific
time period.

Despite of achieving good overall performance in terms of delivery
probability and cost (with a delay increase trade-off), \emph{dLife}
does not reach maximum performance. It is believed that the introduction
of some level of randomness may improve its performance, as randomness
has been proved to increase delivery \cite{people}.

The \emph{Context Information Prediction for Routing in OppNets} (CiPRO)
\cite{cipro} solution also takes advantages of what is happening
in the daily routines of nodes. With \emph{CiPRO}, the message carrier
has enough knowledge about the time and place it will meet other nodes
to forward the message. Such knowledge is defined in terms of node
profile (with the carrier's - name, residence address, workplace,
nationality, ...\textbf{ }- and device's - battery level, memory,
... -\textbf{ }information) which is used to compute\textbf{ }encounter
probability between a node and destination $p_{[N,D]}$ through a
profile match. As in \emph{dLife}, \emph{CiPRO's} encounter probability
can also reflect encounter happening in for specific time periods,
$P_{i}$, which is given by a ratio between the sum of the probability
of all the encountered nodes $E$ by the node towards the destination
$D$, $p_{[E,D]}$, and the set of these nodes, $|N_{i}|.$ When two
nodes meet, \emph{CiPRO} triggers a control message which is propagated
up to its two-hop neighbors and contains the node profile information
of the destination in a evidence/value pair format. Thus, it will
forward messages according to the type of contact, namely occasional
(where neighbors with higher encounter probability - computed based
on the control message - towards the destination will get the message)
and frequent (where the node will use the history of encounters to
a given destination to foresee a specific time period for efficiently
broadcasting control packets and the message itself). For the latter,
a BackPropagation Neural Network model is used to determine the future
encounter probability, $P_{pred}$.

Like \emph{dLife}, \emph{CiPRO} also has a performance trade-off between
delivery probability/cost and delay, and perhaps introducing some
randomness, in the occasional contact case, could bring more improvements
to the solution.

In summary, proposals that take into consideration the idea of social
similarity perform quite well when compared to algorithms simply based
on history of encounters, encounter prediction, and message prioritization.
However, most of the social-aware proposals consider the dynamism
found in node mobility, instead of focusing on the dynamism of the
social information that come from such movements. In addition, Hossmann
et al. (2010) \cite{thyneighbor} show that, if the contact aggregation
considered for determining the proximity graph is based on time window,
some social metrics (e.g., betweenness centrality, similarity) can
lead to node homogeneity regarding the given metric. That is, as network
lifetime increases, these nodes may have the same characteristic (i.e.,
popularity) which will result in great impact on the performance of
forwarding algorithms. This suggests that social-aware solutions must
be carefully designed in order to not end up becoming a mere random-based
solution. Moreover, most of the aforementioned proposals only supports
point-to-point communication. (Multi)point-to-multipoint communication
is a desirable feature in opportunistic routing as it can help reaching
more nodes interested in the content of the messages with better performance
and wise use of resources \cite{multicast}.

Now that we have learned about the social-aware solutions, we next
present the existing taxonomies of opportunistic routing, identifying
the appearance of the social-aware branch, and providing a simplified
taxonomy proposed in 2011{*}{*} with a minor update  given the appearance
of new social-aware opportunistic routing solutions.

\section{Taxonomies\label{sec:Taxonomies}}

The analysis of opportunistic routing approaches shows the existence
of different trends based on distinct goals. On the one hand, we have
single-copy forwarding approaches aiming to optimize the utilization
of network resources. On the other hand, replication-based approaches
aim to optimize delivery probability. Forwarding has the advantage
of using network resources properly, but may end up taking too long
to delivery the messages. Replication-based approaches present a message
delivery probability that is, in almost every solution, very close
to optimal, but with a high cost. Aiming to achieve a good balance
between a high delivery probability and a low utilization of network
resources, several proposals try to avoid flooding the network, by
exploiting mobility of nodes, history of encounters, and social parameters.

Upon what is available in terms of opportunistic routing solutions,
there are different classifications: considering aspects (e.g., level
of knowledge) that led to an unbalanced classification, assigning
most solutions to a few set of categories, or to very specific classification
branches (e.g., by considering for instance information coding or
methods to control movement of nodes). Independently of the classification
approach used, our goal is to provide the reader with an overview
of the existing opportunistic routing taxonomies in order to show
when social similarity started to be considered and that its importance
is recognized as it still appears evident in the latest taxonomies..

\subsection{Existing taxonomies}

The first taxonomy for opportunistic routing was proposed by Jain
et al. (2004) \cite{oracleknowledge} based on three types of classification.
The most important one is the first type of classification, which
divides opportunistic routing according to the knowledge about the
network that nodes need to have to perform message forwarding. The
second and third classifications follow a trend already used to classify
other type of routing: the second approach classifies routing as proactive
(i.e., route computation happens prior to traffic arrival) and reactive
(i.e., route computation takes place upon the need for sending data);
the third approach classifies routing as source-based (i.e., the complete
route is determined by the source), or hop-based (i.e., the next hop
is determined in every traversed hop).

In what concerns the knowledge-based taxonomy, knowledge about the
network is provided by centralized \textit{oracles}. Four different
oracles are proposed: \textit{\emph{i)}}\textit{ Contact Summary Oracle,
}\textit{\emph{which}} provides summarized information about contacts
(e.g., average waiting time until next contact, average number of
contacts);\emph{ }\textit{\emph{ii) }}\textit{Contacts Oracle}\textit{\emph{,
which}} provides more detailed information related to contacts between
nodes at any point in time (e.g., number of contacts in a given period);
\textit{\emph{iii) }}\textit{Queuing Oracle}\textit{\emph{, which}}
provides information about buffer utilization at any time; iv) \textit{Traffic
Demand Oracle, }\textit{\emph{which}} provides information about present
or future traffic demand. According to the authors, the more knowledge
a routing solution can get, the better its performance will be, which
leads to increasing unrealistic approaches since such ubiquitous knowledge
is impossible to get in a dynamic network. 

\begin{figure}
\begin{centering}
\includegraphics[scale=0.7]{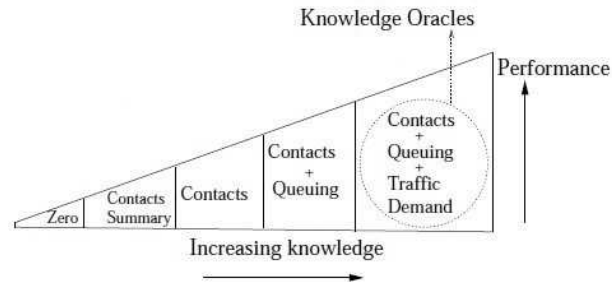}
\par\end{centering}

\caption{\label{fig:oracles}Knowledge oracles (Jain et al. \cite{oracleknowledge})}
\end{figure}

The knowledge levels used by each oracle can be: zero, where solutions
use no information about the network to perform routing; partial,
where solutions can route by using only the \textit{Contact Summary}
oracle, or one/both of the \textit{Contacts} and \textit{Queuing}
oracles; and complete, where all oracles (\textit{Contacts}, \textit{Queuing},
and \textit{Traffic Demand}) are considered. Fig.~\ref{fig:oracles}
illustrates all the different levels of knowledge used in this taxonomy
as well as the existing oracles, and the relationship between performance
and knowledge.

In realistic scenarios existing routing solutions are, however, classified
as having zero or partial knowledge. This is due to the nature of
opportunistic networks where network topology is not known beforehand,
which makes it difficult to have a central entity (e.g., oracle) providing
information to aid routing decisions. Not to mention the delay incurred
to gather/process such information, which can become quite unfeasible
in scenarios with short-lived contacts between nodes. Regarding the
time and place for routing decisions, most of the solutions can be
seen as hop-by-hop reactive approaches.

In our opinion, a taxonomy based on network knowledge, as well as
proactive/reactive and source/hop characteristics is not the most
suitable one to provide a balanced classification of opportunistic
routing approaches.

The most complete taxonomy for opportunistic routing, to our knowledge,
is provided by Zhang (2006) \cite{zhang06}. This taxonomy follows
the work presented by Jain et al., classifying protocols according
to the information they require from network as well as relevant routing
strategies (e.g., pure forwarding, estimation of forwarding probability).
The result is a taxonomy based on two categories, namely deterministic
and stochastic routing. In the former case, node movement and future
connections are known beforehand (i.e., nodes are completely aware
about topology), which follows the oracle-based taxonomy proposed
by Jain et al. In the case of stochastic approaches, the behavior
of nodes and network is random and unknown. Hence, routing decisions
depend upon local conditions leading to simple solutions where messages
are replicated on every contact up to more complex solutions where
the use of history of encounters, node mobility patterns, and message
coding are considered for routing decisions. Zhang also classifies
proposals regarding whether (or not) node movement can be controlled.

We can say that Zhang's proposal complements the proposal of Jain
et al. by including a set of more realistic (stochastic) approaches.
However, Zhang's proposal emphasizes aspects that are orthogonal to
different routing categories (i.e., coding methods) and puts emphasis
on categories that are specific to deterministic networks (e.g., node
movement control), which do not represent a scenario where opportunistic
routing may have greater impact in our daily life.

Balasubramanian et al. (2007) \cite{rapid} propose a taxonomy based
on two types of classification criteria regarding the routing strategy
and the effect on performance metrics. Based on the first criterion,
solutions are divided into two routing strategies: replication-based,
where messages are replicated and then transferred to the next hop;
and, ii) forwarding-based, where only one copy of the message traverses
the network. The second classification criterion is used to divide
solutions based on the effect routing decisions have on performance
metrics. There are two types of effect: i) incidental, where the effect
of decisions do not take into consideration resource constraints;
and, ii) intentional, which determines the effect of a routing strategy
on metrics considering such constraints.

We believe that classifying routing solutions according to their routing
strategy is the most suitable approach. However, there is more to
it than what is presented by Balasubramanian et al. since there are
other performance metrics, besides incidental/intentional usage of
resources, that can be used to distinguish among solutions that follow
the same generic strategy (i.e., replication). 

The same idea of classifying solutions based on their routing strategy
(i.e., forwarding/replication) is also followed by Song and Kotz (2007)
\cite{prediction} and Nelson et al. (2009) \cite{ebr}. In relation
to the work presented by Balasubramanian et al., the classification
presented by Song and Kotz is able to further divide replication-based
proposals taking into consideration, not only the effect that they
have on network resource consumption, but also on delivery probability.
Still in what concerns the classification of replication-based approaches,
Nelson et al. propose to divide them into flooding-based and quota-based.
This classification is quite important since its shows that it is
not the fact that messages are duplicated that may lead to a flooding
situation, as the one that occurs with epidemic routing, but the employed
routing metric for deciding on replication. Nelson et al. shows that
there are probabilistic-based replication strategies that actually
end up flooding the network (flooding-based), since at the end of
the experimental period every node has at least one replica of each
message, while others (quota-based) have higher success in controlling
the number of replicas in the network. However, this taxonomy is incomplete
in the sense that it does not consider routing categories based on
metrics such as encounter number and resource usage.

In another survey \cite{survey2010}, D'Souza and Jose (2010) classify
routing solutions into three major categories: i) flooding-based,
where nodes flood the network to increase delivery probability or
apply some measures to control such flooding by bounding the number
of messages copies to be distributed in the network and by embedding
additional information to messages blocks to reduce flooding effects;
ii) history-based, in which the history of encounters between nodes
is taken into account to improve routing decisions; and, iii) special
devices-based, where stationary or mobile devices are used to improve
communication among the communicating nodes. Such special devices
can also consider social interaction among nodes to perform routing
decisions. Like Zhang, D'Souza and Jose consider aspects that are
orthogonal (e.g., network and erasure coding) and could be easily
applied to other categories. Despite the fact that the first social-aware
solution appeared in 2007 (i.e., Label \cite{label}), such category
of opportunistic routing only made it to taxonomies in 2010 with D'Souza
and Jose's work. Still, their taxonomy proposal includes this trend
under a category (i.e., special devices-based) that does not comply
with the regular behavior (i.e., random and unknown) found in opportunistic
networks. We believe that distinguishing proposals according to whether
or not they use special stationary/mobile devices to improve data
exchange is not realistic as the network/nodes will have to present
a deterministic behavior in order to correctly place these devices
in the system.

The classification proposed by Spyropoulos et al. (2010) \cite{tax2010}
groups opportunistic routing proposals according to message exchange
scheme they employ: forwarding (only one message copy traverse the
network), replication (message is replicated in different levels ranging
from every node getting a copy up to more elaborate solutions based
on utility functions), and coding (where message can be coded and
processed at the source or as it travels throughout the network).
The authors also identify the different types of utility functions
that can be applied to either forwarding or replication message schemes.
Such functions are categorized according to their dependency on the
destination (i.e., destination dependent/independent). Additionally,
they classify DTNs according to characteristics that have major impact
on routing such as connectivity, mobility, node resources, and application
requirements. The authors succeed in mapping the routing solutions
to the different types of DTNs. Still, the proposed opportunistic
routing classification considers categories that can be orthogonal
(i.e., coding) and does not include the social similarity trend observed
in 2007. They simply refer to the social aspects as a mere destination-dependent
function in which we believe comprises a new research direction that
includes social relationships, interests, and popularity to achieve
suitable delivery probability with shorter delay and cost.

\begin{figure}
\begin{centering}
\includegraphics[scale=0.4]{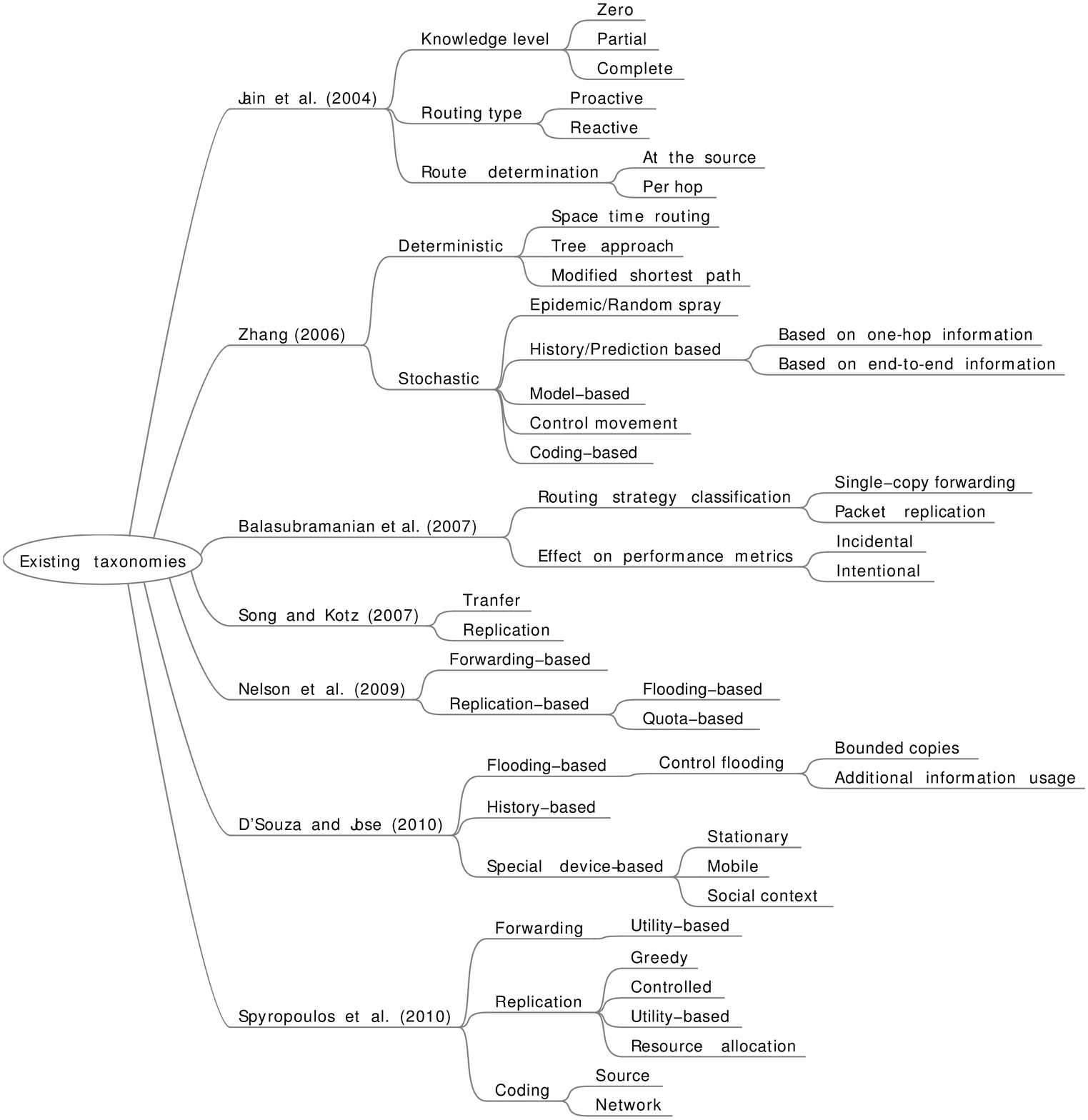}
\par\end{centering}

\caption{\label{fig:taxonomies}Existing taxonomies}
\end{figure}

Fig.~\ref{fig:taxonomies} summarizes the analyzed taxonomies. Such
figure shows an evolution towards stochastic approaches that do not
require any knowledge about the global network topology. Deterministic
approaches, based on some kind of centralized oracles, are not realistic
and do not reflect the behavior found in opportunistic networks. Within
stochastic approaches, and since 2007, there is a clear trend to classify
routing strategies considering their success in achieving a good balance
between delivery probability (e.g., message replication) and usage
of network resources (e.g., forwarding) by employing the social similarity
approach. 

Keeping in mind the ultimate goal of balancing performance and resource
usage, current taxonomies are not capable of illustrating the different
families of routing metrics that have been used to devise replication-based
approaches able to avoid network flooding. Next, we present a simplified
proposal to extend the taxonomy (presented by Nelson et al.) that
most reflects the behavior of opportunistic networks with a set of
categories that represent recent trends in stochastic opportunistic
routing.

\subsection{Proposed Taxonomy}

One can conclude that the existing taxonomies generally focus on the
analysis of opportunistic routing proposals based on their efficiency
(e.g., level of knowledge employed to achieve higher delivery rates
\cite{oracleknowledge}, forwarding schemes that result in different
performance levels \cite{zhang06,rapid,prediction}, limiting the
number of messages copies in the network to spare resources \cite{ebr,survey2010},
or application of the correct routing algorithm according to the specificity
of the network \cite{tax2010}), but they lack an analysis of the
characteristics of the graph structure used.

We believe that focusing on an analysis of the topological features
(e.g., contact frequency and age, resource utilization, community
formation, common interests, node popularity) assumed by each proposal
may lead to a more stable taxonomy useful to study real-world networks
such as computer networks operated based on social behavior.

Thus, in this section, we present a 2011 taxonomy including updates
based on the emergence of recent opportunistic routing proposals (cf.
section \ref{sec:Social-aware-Opportunistic-Routi}). 

Social similarity is an example of recent metrics that have clearly
created a new trend in the investigation of opportunistic routing
\cite{label,simbet,socialcast,bubblerap,people,dlife,cipro}. The
reason for this is that social behavior takes into account human relationship
characteristics such as contacts with other people, time spent with
these people, the level of relationship between people, among others.
And, since computing devices are carried by humans, social-based forwarding
decisions can consider people's socially meaningful relationships,
where the relevant information come from aspects such as human mobility,
interaction and social structures. This information can be used to
perform forwarding, because the topology created from human social
behavior varies less than the one based on mobility and thus such
solutions deserve being categorized.

It is important to mention that all studied opportunistic routing
proposals take advantage of node mobility to forward data ahead, where
some of them (e.g., \emph{Epidemic} \cite{epidemic}, \emph{Direct
Transmission} \cite{single-copy-fwd1}) are rather simple and use
the resulting contacts to reach the destination, while others are
more elaborate and consider social aspects in order to find the destination
(e.g., \emph{LABEL} \cite{label}, \emph{PeopleRank} \cite{people}).
This is the reason why Moreira et al. (2011) do not devote a specific
category (as \emph{Model}- or \emph{Control Movement-based }in Zhang
\cite{zhang06} and \emph{Mobile Device-based} in D'Souza and Jose
\cite{survey2010}) since this is an inherent feature of opportunistic
proposals and instead look for features that help understanding their
graph structure.

The proposed taxonomy is based on an initial classification of all
proposals as forwarding-, flooding-, or replication-based. The forwarding-based
category is also known as single-copy forwarding (e.g., \emph{MEED},
and approaches in Spyropoulos et al. \cite{single-copy-fwd1}) since
all approaches propose that only one copy of each message traverses
the network towards the destination. From the resource consumption
viewpoint, this category of approaches is quite interesting since
it keeps network (e.g., bandwidth) and node (e.g., buffer space) resources
usage at a low level; however, all approaches suffers in general from
high delay rates that, consequently, results in a low delivery rate. 

Nelson et al. propose, depending on the level of duplication, that
algorithms within the replication-based approach be divided into flooding-based
and quota-based. The flooding-based algorithms are able to increase
delivery rate to a very high level, whereas the quota-based algorithms,
in general, allow a more wise usage of resources, resulting in low
delay and reduced flooding overhead since they tend to spread less
copies of messages in the network. 

Moreira et al. do consider these different levels of replication but
unlike Nelson et al., it is proposed that flooding-based algorithms
be classified out of the replication branch. First, because only proposals
that allow every node to spread a copy of each message to every other
node that they meet (e.g., \emph{Epidemic}) are considered. And, also
due to the fact that having (or not) the quota-based feature (i.e.,
where the number of created copies does not depend on the number of
network nodes) can be found in the different algorithms identified
in Section \ref{sec:Social-oblivious-Opport}.

Despite being an aggressive approach, the flooding-based strategy
is able to increase delivery rate, but at the same time leads to a
high consumption of resources. Such waste of resources can be avoided
with algorithms that try to somehow control flooding. This control
starts by limiting the number of copies injected in the network, if
it is able to avoid nodes ending up with a replica of every created
message, the algorithm has the quota-based feature.

So, the replication-based approaches have as common goal an attempt
to increase the delivery rate by sending several copies of the initial
message through different nodes to quickly reach the destination before
message expiration time. Since these approaches consider different
routing algorithms and metrics, these are its sub-categories. 

The first sub-category is the encounter-based, where nodes choose
next hops based either on frequency encounters (e.g., \emph{PROPHET},
\emph{MaxProp}, and \emph{Prediction,} and \emph{EBR}), or aging encounters
(e.g., \emph{FRESH}, \emph{EASE}, and \emph{Spray and Focus}). In
the former, proposals consider the history of encounters with a specific
destination to support opportunistic forwarding of messages or the
frequency nodes met in the past to predict future encounters. As for
the latter, proposals consider the time elapsed since the last encounter
with the destination to decide about next hops.

Resource usage is the second sub-category, in which decisions are
made considering the age of messages (e.g., \emph{Spray and Wait},
and \emph{OPF}) or knowledge about local resources (e.g., \emph{PREP},
and \emph{RAPID}). Aging messages proposals have in common the fact
that they aim to avoid messages to be kept being forwarded in the
network by creating metrics that define the age of message copies.
As for the resource allocation proposals, they take forwarding decisions
that wisely use available resources.

The last sub-category is related to social similarity, where proposals
start to follow more complex algorithms aiming first at avoiding flooding
with high probability, and exploiting social behavior. Thus, social
similarity algorithms are divided into: community detection, shared
interest, and node popularity.

Community detection approaches (e.g., \emph{SimBet}, \emph{Label,
BubbleRap}) rely on the creation of node communities taking into consideration
people social relationships translated to contact numbers and duration
of contact among nodes. These approaches suffer with the overhead
of community formation. The shared interest approach (e.g., \emph{SocialCast})
relies on the assumption that nodes with the same interest as the
destination of the message are good forwarders since they have high
probability to meet. But this assumption may not always be true \cite{people},
since a node with similar interest to a given group may not even come
in contact with this group of nodes. Still within the social similarity
category, there are approaches that are based only on a process of
ranking people in terms of their popularity without a straight dependency
upon neither the computation of communities nor the synchronization
of interests. Node popularity approaches (e.g., \emph{PeopleRank})
make use of social information\emph{ }to generate ranks to nodes based
on their position on a social graph, using such ranking to decide
on the next hop for data exchange. Although social similarity algorithms
provide stable graphs, it is proven that relying on socially connected
nodes may not be enough to guarantee a good performance, which can
be improved with the inclusion of some degree of randomness in the
forwarding decision \cite{people}.

\begin{figure}
\begin{centering}
\includegraphics[scale=0.55]{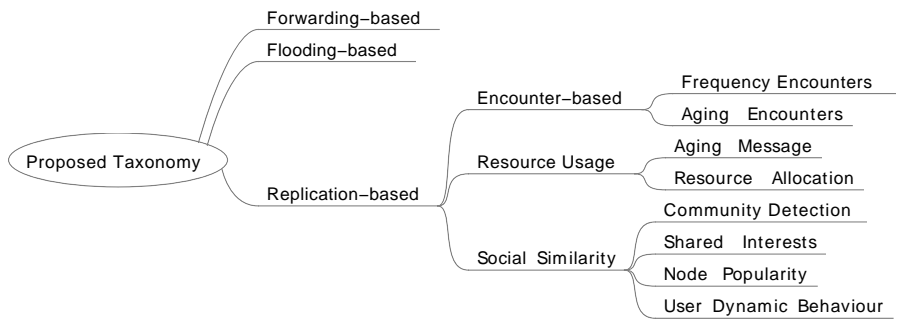}
\par\end{centering}

\caption{\label{fig:myTaxonomy-1}Taxonomy for opportunistic routing}
\end{figure}

With the appearance of opportunistic routing solutions which take
into account the dynamism of user social behavior (i.e., \emph{dLife}
and \emph{CiPRO}), we propose an update to the taxonomy of Moreira
et al. So, Fig.~\ref{fig:myTaxonomy-1} illustrates the proposed
taxonomy that complements the most recent trend (replication vs. forwarding)
with the analysis done of twenty-two proposals published between 2000
and 2012 and includes the new sub-category, user dynamic behavior.

It is easily observed that all categories of the presented taxonomy
has advantages and disadvantages. However, our goal was not to identify
the winner category, but to: i) present the reader with reasons to
support the emergence of the new trend based on social similarity;
and ii) update a previously proposed taxonomy with a new sub-category
based on the latest social-aware opportunistic routing proposals.

\section{Experiments\label{sec:Experiments}}

We start this section by describing the employed evaluation methodology
and simulation settings. Then, we present our considerations on the
obtained results pointing out the advantages and constraints of using
social similarity existing among nodes to perform forwarding in opportunistic
networks.

The results are presented in two parts: first, we show the results
in a scenario with synthetic mobility models (hereafter referred to
as heterogeneous scenario); then, we present a performance comparison
between some of the aforementioned opportunistic routing proposals
to support our case: social similarity is indeed a trend and has great
potential in improving opportunistic routing\emph{.}

\subsection{Evaluation Methodology}

Our experiments are done based on the Opportunistic Network Environment
(ONE) simulator \cite{one}.

For that, we run simulations representing a 12-day interaction period
(with 2 days of warmup, not considered for the results) in our heterogeneous
scenario. For the human trace-based simulations, we considered the
Cambridge traces \cite{cambridge-haggle-imote-content-2006-09-15}.
Each simulation is run ten times (with different random number generator
seeds for the used movement models) in order to provide results with
a 95\% confidence interval. All results are analyzed considering the
average delivery probability (i.e., ratio between the number of delivered
messages and total number of created messages), average cost (i.e.,
number of replicas per delivered message), and average latency (i.e.,
time elapsed between message creation and delivery).

Regarding the ONE simulator, the time step size is of 2 seconds for
the heterogeneous scenario and of 1 second for the trace-based scenario
for more precision in the collected data. All simulations are performed
in batch mode with 2 GB RAM dedicated memory.

\subsection{Simulation Settings \label{sub:Settings}}

With our heterogeneous scenario simulation settings, we aimed at creating
a scenario that is as close as possible to the heterogeneous environment
(in terms of mobility) that people find in their daily activities
in a city, such as pedestrian walks, combined with bus and car rides,
as well as people attraction to diverse locations such as work, shopping
places and home.

Our simulation scenario is part of the Helsinki city (available in
ONE) and has 150 nodes distributed in 17 groups (8 groups of people
and 9 groups of vehicles). All nodes are equipped with one WiFi interface
(11 Mbps/100 m). One of the vehicle groups, with 10 nodes, follows
the \emph{Shortest Path Map Based Movement} (SPMBM) mobility model,
based on which nodes randomly choose a point in the map and use the
shortest path to reach it. These nodes represent for instance police
patrols. They move with speed between 7 to 10 m/s and have a waiting
time between 100 and 300 seconds when arriving at the chosen destination.

The other eight vehicle groups represent buses that cover different
parts of the city. Each group is composed of 2 vehicles each. They
follow the \emph{Bus Movement} mobility model with speeds between
7 to 10 m/s and have waiting times between 10 and 30 seconds.

Regarding groups of people, they follow the \emph{Working Day Movement}
(WDM) mobility model with walking speeds ranging from 0.8 to 1.4 m/s.
People may also use buses to move around the city. Each group has
different meeting spots, offices, and home locations. People spend
8 hours at work and present 50\% probability of having an evening
activity after leaving work. In the office, nodes move around and
have a pause time ranging from 1 minute to 4 hours. Evening activities
can be done alone or in a group, with a maximum of 3 people each.
Each evening activity can last between 1 and 2 hours. 

For the human trace-based simulations, the Cambridge trace was considered
which corresponds to a 2-month imote communication between 36 students
carrying these devices throughout their daily activities. 

The traffic load used in the simulations comes from a file previously
generated, as has established source/destination pairs, where approximately
500 messages are generated per day among a subset of node pairs. In
a simulation of 12 days (i.e., heterogeneous scenario), with 2 days
of warmup time, this results in a total of 6000 messages, from which
5018 messages are considered for the performance assessment. As for
the trace-based simulations, all the 6000 messages are considered
in the assessment.

Message TTL values are set at either 1, 2 and 4 days, as well as 1
and 3 weeks. Since we want to bring our experiments as close as possible
to the real world, we chose values that can represent the different
applications which cope with opportunistic routing. Message size ranges
from 1 kB to 100 kB. The buffer space is of 2 MB as users may not
be willing to share all of their storage space. Message and buffer
size comply with the universal evaluation framework that we proposed
previously \cite{survey,latincom,ieeeLA} based on the evidence that
prior-art on opportunistic routing (19 proposals from 2000 to 2010)
follows completely different evaluation settings, making the assessment
a challenging task.

The proposals considered for our experiments comprise social-oblivious
as well as social aware solutions. As representative of the former
class of opportunistic routing, we chose \emph{Epidemic} \cite{epidemic}
(normally serves as upper bound for delivery probability), \emph{PROPHET}
v1 \cite{prophet} (accepted as standard in by the Delay Tolerant
Networks research community), and \textit{Spray and Wait }\textit{\emph{\cite{spraywait}}}
(appearing as lower bound for latency) for being the most cited proposals
(i.e., often used as benchmark for performance comparison) \cite{survey}.
For the social-aware approaches, \emph{Bubble Rap} \cite{bubble2011},
\emph{dLife and dLifecomm} \cite{dlife} were selected as representative
of solutions considering social structures (i.e., communities), node
popularity, and dynamic behavior of users.

Some of the simulated proposals need to have some parameters set,
thus we set: i) \emph{PROPHET} with aging of delivery predictability
happening at every 30 seconds; ii) \textit{Spray and Wait}, being
binary and with a number of spraying copy set to 10; iii) \emph{dLifeComm}
and Bubble \emph{Rap}, with K-Clique ($k=5$, 700-second familiar
threshold) and cumulative window algorithms for community formation
and node centrality computation; and iv) \emph{dLife} and \emph{dLifeComm},
with 24 daily samples. The choice for such values are among those
which each proposal has presented good overall performance as specified
in their respective original papers.

Table \ref{tab:Simulation-parameters} summarizes the setup parameters.

\begin{table}
\caption{\label{tab:Simulation-parameters}Simulation parameters}

\begin{tabular}{p{2.3cm}p{2cm}p{2cm}p{2cm}p{1cm}}
\hline 
\noalign{\smallskip{}
} Parameters  & Values &  &  & \tabularnewline
\noalign{\smallskip{}
}\svhline\noalign{\smallskip{}
} Simulator  & \multicolumn{4}{l}{Opportunistic Network Environment (ONE)}\tabularnewline
Routing Proposals  & \multicolumn{4}{l}{Epidemic, PROPHET, Spray\&Wait, Bubble Rap, dLife and dLifeComm}\tabularnewline
Scenarios  & \multicolumn{2}{l}{Heterogeneous} & \multicolumn{2}{l}{\quad{}Trace Cambridge}\tabularnewline
Simulation Time  & \multicolumn{2}{l}{1036800 sec} & \multicolumn{2}{l}{\quad{}1000000 sec}\tabularnewline
\# of nodes  & \multicolumn{2}{l}{150 (people/vehicles)} & \multicolumn{2}{l}{\quad{}36 (people)}\tabularnewline
Mobility Models  & \multicolumn{2}{l}{Working Day, Bus, Shortest Path Map Based} & \multicolumn{2}{l}{\quad{}Human}\tabularnewline
Node Interface  & \multicolumn{2}{l}{Wi-Fi (Rate: 11 Mbps / Range: 100 m)} & \multicolumn{2}{l}{\quad{}Bluetooth}\tabularnewline
Node Buffer  & \multicolumn{4}{l}{2 MB}\tabularnewline
Message TTL  & \multicolumn{4}{l}{1, 2, 4 days, 1 and 3 weeks}\tabularnewline
Message Size  & \multicolumn{4}{l}{1 \textendash{} 100 kB}\tabularnewline
\# of Messages  & \multicolumn{4}{l}{6000 (only 5018 considered for the heterogeneous scenario)}\tabularnewline
Spraying copies  & \multicolumn{4}{l}{L = 10 (Spray and Wait)}\tabularnewline
K-Clique  & \multicolumn{4}{l}{k = 5 and familiarThreshold = 700 seconds (Bubble Rap and dLifeComm)}\tabularnewline
Daily Samples  & \multicolumn{4}{l}{24 (dLife and dLifeComm)}\tabularnewline
\noalign{\smallskip{}
}\hline\noalign{\smallskip{}
}  &  &  &  & \tabularnewline
\end{tabular}
\end{table}

It is important to say that our goal is not to show which proposal
is the best. Instead, we want the reader to understand the potential
of social-aware opportunistic solutions and why they haven shown to
be a trend in the last years for opportunistic networks.

Next we present the experiments results.

\subsection{Results}

Before we present the results of our experiments, here are a few general
observations regarding our findings. The average number of contacts
per hour is of approximately 962 in the heterogeneous scenario and
of 32 in the trace-based one. Additionally, contacts are more sporadic
in the trace-based scenario than in the heterogeneous one, in which
contact frequency is more homogeneous. We also observe that the average
number of unique communities is higher in the heterogeneous scenario
(\textasciitilde{}68) than in the trace-based scenario (\textasciitilde{}8.7).
Furthermore, most of the created communities encompasses all the existing
nodes (150 for the heterogeneous simulations, and 36 for trace), which
means that independently of the level of contact homogeneity, nodes
are still well connected. Heterogeneous scenario

\begin{figure}[H]
\begin{centering}
\includegraphics{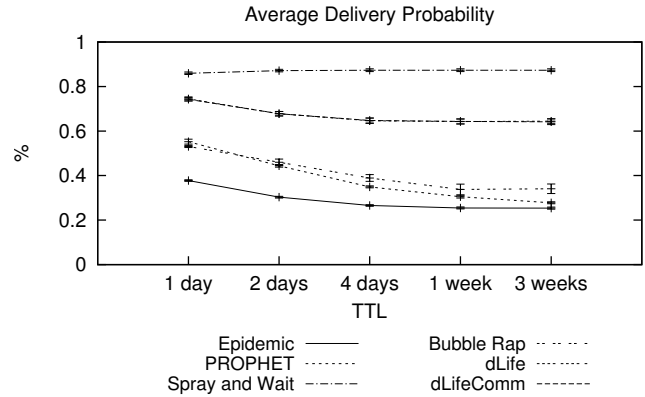}
\par\end{centering}

\caption{\label{fig:adp_sce}Average delivery probability}
\end{figure}

Fig. \ref{fig:adp_sce} presents the results for the average delivery
probability. It is clear that \emph{Epidemic} (normally seen as an
upper bound for this performance metric) has the worst performance
amongst the proposals. This is due to the available buffer space (2
MB) which was reduced as to represent a limited willingness of the
user to carry information of behalf of others. 

As we could see \cite{survey,latincom,ieeeLA}, authors tend to consider
unlimited storage. However, while this may be true in scenarios comprising
nodes for the serving purpose (e.g., \cite{rapid}), normally a user
may find him/herself in an opportunistic network formed on-the-fly
and surrounded by other users sharing storage space according to their
devices' capabilities and mostly to their willingness in cooperating. 

\emph{PROPHET} and \emph{Bubble Rap} are also affected by the limited
buffer and their delivery capability diminishes as TTL increases.
Since messages are allowed longer in the network (i.e., being replicated),
this consequently results in buffer exhaustion. We believe that the
mobility heterogeneity found in the scenario also contributes for
the decrease in the delivery probability.

Since the dynamic behavior of users are considered, both \emph{dLife}
and \emph{dLifeComm} are less affected by the limited buffer space.
They carefully decide whether to replicate based on the social weight
towards the destination and node importance in the network at the
time of the encounter. Thus, wiser decisions take place as next forwarders
are only chosen if they indeed have a stronger social connection or
importance than the current message carriers.

\emph{Spray and Wait }has the best performance. This is due to the
fact that the scenario comprises nodes (e.g., buses and police patrols)
which cover most of the simulated area and this proposal takes advantage
of that. Most of its random replications happen to be to such nodes
and as they move across the entire scenario added to the 100-meter
transmission range, its delivery capability increases. Despite the
better performance, \emph{Spray and Wait }still does not take advantage
of longer TTLs to reach optimum delivery as some of the messages end
up in the possession of nodes which are not well socially connected
to their destinations.

\begin{figure}[H]
\begin{centering}
\includegraphics{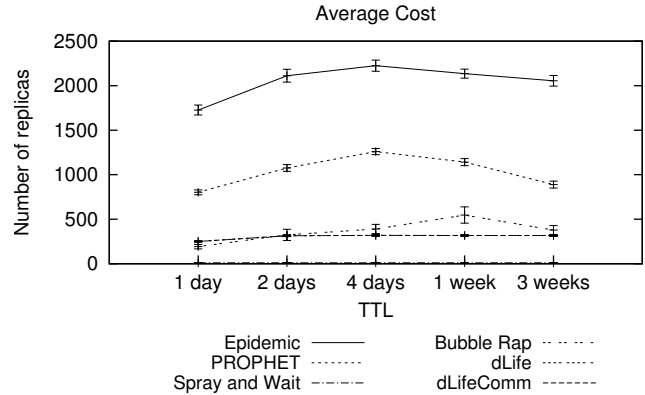}
\par\end{centering}

\caption{\label{fig:ac_sce}Average cost}
\end{figure}

Fig. \ref{fig:ac_sce} shows the performance of each proposal considering
the number of replicas created per delivered message. As one could
expect, \emph{Epidemic} and \emph{Spray and Wait} are the upper and
lower bounds for this metric. The former has the highest cost as it
replicates a message with every encountered node which does not have
a copy yet. The scenario itself is not epidemic-friendly since it
has a high number of contacts among nodes. On the other hand, \emph{Spray
and Wait} is allowed to create up to ten ($L=10$) copies per messages,
and thus its cost is the lowest (average across simulations of \textasciitilde{}10.16
replicas per delivery). It is important to note that the way the ONE
simulator determines cost (a.k.a. overhead) is given by the ratio
between the number of successful relayed message (discarding the relaying
to the final destination, i.e., number of delivered messages) and
number of delivered messages. This explains why cost here exceeds
the one \emph{L} specifies for \emph{Spray and Wait}.

By being a probabilistic-based solution, PROPHET considers the frequency
of past contacts with the message's destination to decide on replication.
Since the proposal ages such encounters, only nodes that have frequent
contacts with the destination will be entitled to receive a copy.
Despite the effort of the proposal, it still requires a high number
of replicas to perform a successful delivery.

\emph{Bubble} \emph{Rap}, \emph{dLife} and \emph{dLifeComm} have a
much lower cost when compared to \emph{Epidemic} and \emph{PROPHET},
which shows that taking forwarding decisions based on some level of
social similarity is beneficial as messages are replicated to nodes
well socially connected to the destination of message. Regarding \emph{Bubble
Rap}, its cost is expected to increase as TTL increases \cite{bubblerap,bubble2011};
however, there is still a drop in cost observed for a 3-week TTL.
We believe this is due to the fact that the proposal has taken advantage
of the longer TTL, choosing next forwarders much more wisely, which
reduces its cost. As a matter of fact, this can be confirmed in Fig.
\ref{fig:adp_sce} where the same performance is achieved but with
much less replications.

Both \emph{dLife} and \emph{dLifeComm} present a much more stable
behavior as they consider the dynamism found in the users' daily routines
and can choose to replicate only to nodes that actually have the best
social interactions with the message's destination at each encounter. 

\begin{figure}[H]
\begin{centering}
\includegraphics{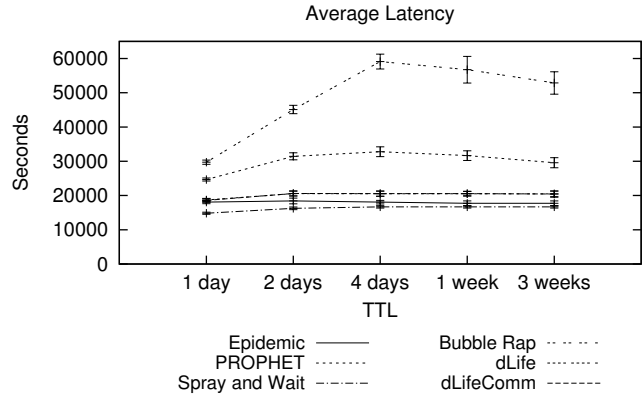}
\par\end{centering}

\caption{\label{fig:al_sce}Average latency}
\end{figure}

A Fig. \ref{fig:al_sce} shows the performance of the proposals in
terms of average latency for delivering messages. As mentioned earlier,
\emph{Spray and Wait} takes advantage of the scenario which has nodes
(i.e., buses and police patrols) covering the whole extension of the
scenario and with longer transmission range. This significantly reduces
this proposal's latency as destinations can be reached in shorter
periods of time as reported in its original paper \cite{spraywait}.
\emph{Epidemic} also has short latencies as spreading many copies
of the same message also reduces the time it will take to reach its
destination.

As waiting for the best (i.e., socially speaking) next forwarder may
take some time, \emph{dLife} and \emph{dLifeComm} have a slight increase
(varying between 490 and 4350 seconds) in latency when compared to
both \emph{Epidemic} and \emph{Spray and Wait}. Thus, one can observe
the tradeoff when only forwarding messages in the presence of strong
social links or highly important nodes in the current daily sample.
Still, despite of taking a little longer reach their destinations,
\emph{dLife} and \emph{dLifeComm }are able to deliver a reasonable
amount of messages. 

\emph{PROPHET} does not consider the social strength of the next forwarder
with the destination; instead, it looks at the frequency of past interactions
and this results in replication of messages that may take longer to
reach their destinations. This can also be confirmed with the performance
of \emph{Bubble} \emph{Rap} regarding this metric: it chooses also
to replicate according to the centrality (i.e., cumulative number
of unique contacts within 6-hour time windows) of nodes. Centrality
indeed identifies nodes which may be well connected throughout the
the entire experiments, but fails to capture the different levels
of ``being socially connected)'' during the different periods of
the day. To make matters worst given the number of average contacts
per hour in this scenario (962), this proposal rely solely on global
centrality while communities are forming. The result is replicas being
created to nodes with weak social ties to destinations and which account
for the total average latency experienced to deliver messages.

After looking at the performance of the proposals in a scenario of
synthetic mobility models, we next present the performance of the
same proposals using real human traces. With this, we expect to have
experiments which are as close as possible from the reality when it
comes to human mobility.

\subsubsection{Cambridge scenario}

\begin{figure}[H]
\begin{centering}
\includegraphics{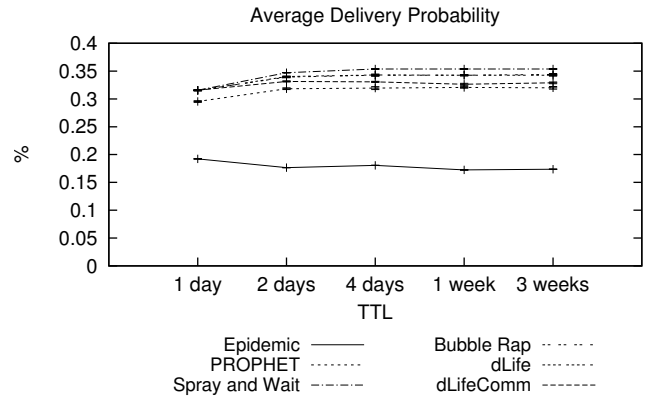}
\par\end{centering}

\caption{\label{fig:adp-trace}Average delivery probability}
\end{figure}

Fig. \ref{fig:adp-trace} presents the performance of the considered
proposal in terms of average delivery probability. One can easily
see that the advantage of \emph{Spray and Wait} seen in the previous
scenario has decreased. This is mainly due to the fact that in this
scenario there are no nodes covering the whole area. Instead, nodes
are encountering others as they move throughout their daily routines.
Additionally, contacts are much more sporadic. Under these circumstances,
\emph{Spray and Wait} spreads copies to nodes that may never come
in contact with the destination, thus reducing its delivery capability.

This contact sporadicity also affects the remaining proposals since
the solutions somehow depend on how nodes interact among themselves.
Creating communities and determining centrality and importance of
nodes as well as social weight among them take much longer in a scenario
with low number of contacts (32 per hour). This is why \emph{dLife}
and \emph{Bubble} \emph{Rap} are statistically equivalent and perform
very much the same. \emph{dLifeComm} relies on the node importance
to replicate, as node importance is more elaborate and will take longer
to distinguish really important nodes, message will be replicated
to nodes not well socially related with the destination. Despite being
based in the notion of community formation, \emph{Bubble} \emph{Rap's}
centrality is simple and can distinguish nodes much easier.

\emph{PROPHET} faces the same issue, delivery predictabilities of
nodes are not well computed and nodes getting copies are not the best
option to increase its delivery capacity. And despite of the uncontrolled
replication, \emph{Epidemic} does not reach its best in this scenario
as nodes meet occasionally added to limited buffer.

\begin{figure}[H]
\begin{centering}
\includegraphics{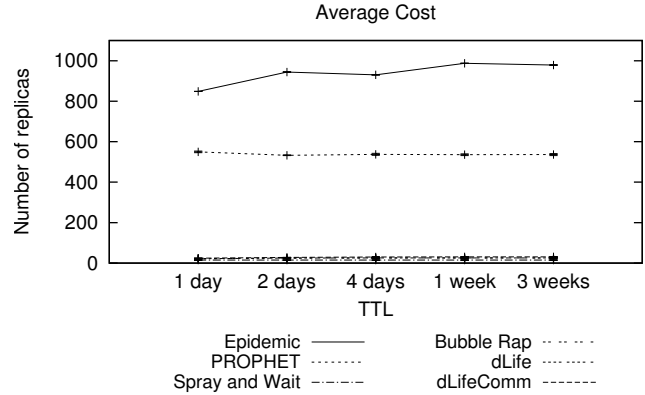}
\par\end{centering}

\caption{\label{fig:ac-trace}Average cost}
\end{figure}

In Fig. \ref{fig:ac-trace} we can see also the effect of the sporadic
contacts. The number of copies to perform a delivery is much less
since there are only few nodes to receive such copies at the time
of exchange. Still \emph{Epidemic} is the proposal that most replicates
(despite of a decrease varying between 877 and 1300 replicas per delivery)
and \emph{Spray and Wait} remains the proposal with the least cost,
as one could already expect; however, this time with a little increase
(average across simulations of \textasciitilde{}15.12 replicas per
delivery).

\emph{PROPHET }experienced reductions varying between 252 and 606
replicas per delivery and remains as the second proposal to consume
more resources. The social-aware solutions, namely \emph{Bubble} \emph{Rap},
\emph{dLife} and \emph{dLifeComm}, managed to have\emph{ }an average
across simulations of approximately 24.52, 24.56, and 28.79 replicas
per delivery. Here we can observe the potential of social-aware opportunistic
routing as with a few extra copies they can almost reach the same
delivery of the social-oblivious \emph{Spray and Wait} proposal.

\begin{figure}[H]
\begin{centering}
\includegraphics{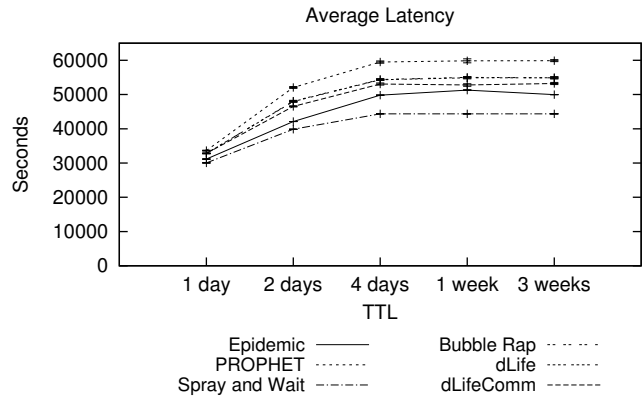}
\par\end{centering}

\caption{\label{fig:al-trace}Average latency}
\end{figure}

As contacts are rather sporadic, one could expect an increase in latency
when delivering messages in the trace-based scenario. Indeed, this
can be observed in Fig. \ref{fig:al-trace} where all the proposals
took longer times to perform deliveries when compared to the heterogeneous
scenario. To influence even more in the time to deliver messages,
we observe an increase in the distance (i.e., average number of hops
of at most 2) to reach the destination in the trace-based simulations.
This means that, besides the time taken to decide whether or not to
replicate, delivery also accounts for the time messages traverse different
hops. The consequence is that for all cases, proposals almost doubled
their time to deliver messages.

We believe this is due to the fact that most of the nodes encounter
one another (although in a sporadic manner) through the simulations,
which leads proposals to not have information reliable enough to suitably
decide on replication. Thus, an increase in the distance to reach
destinations.

Amongst the proposals \emph{Bubble Rap} was the proposal that least
experienced an increase in latency, having an almost similar behavior
to the one observed in the heterogeneous scenario. As a matter of
fact, this proposal experienced a decrease in latency with TTL of
4 days (less 4709 seconds) and 1 week (less 1725 seconds). This scenario
has not been favorable for \emph{PROPHET} (with the highest latency
values), as contact sporadicity really affects the notion of time
of last encounter employ by the proposal. \emph{dLife} and \emph{dLifeComm
}were affected in the sense that the scenario is not dynamic given
the non-homogeneous frequency of contacts among nodes. Since it spreads
many copies, \emph{Epidemic} still keeps its position of second best
solution in terms of latency as one of these many replicas will reach
the destination in short times. And \emph{Spray and Wait} has the
overall latency when it comes to delivering messages in shorter times.

\subsubsection{Considerations}

A subset of social-oblivious and social aware solutions are evaluated
under the same conditions and considering two different scenario.
Each of these scenarios have particularities and specific challenges.
In the first, heterogeneous scenario, nodes follow different mobility
models (representing people, buses and police patrols) and interact
with one another as the move between home and work, go to a leisure
activity, transport people and secure the city. This is a scenario
with a high numbers of contacts among nodes where social similarity
can be easily inferred as they move throughout their daily routines. 

We could observe that this scenario is advantageous to the social-oblivious
\emph{Spray and Wait} as it comprises nodes that cover most of the
simulated area and by randomly spraying copies to such nodes, \emph{Spray
and Wait }manages to fast deliver the highest number of messages (\textasciitilde{}86\%)
with the lowest cost (at most 10.12 copies to perform a delivery).
Still, the social-aware solutions, \emph{dLife} and \emph{dLifeComm},
are able to capture the dynamicity of user behavior and with a small
latency tradeoff (25\% increase when compared to \emph{Spray and Wait})
they reach up to 74\% delivery with a wise usage of resources (i.e.,
buffer space) producing at most 319 replicas to perform a successful
delivery.

Since user dynamicity is not considered, \emph{Bubble Rap} suffers
from the problem of forming communities: as they are not readily available,
the proposal will solely rely on centrality to replicate information,
which also take some time to represent the system's reality. This
results in a delivery probability not more than 53\% with cost reaching
up to \textasciitilde{}547 replicas to successfully deliver messages
and latency increase up to 250\% (in relation to \emph{Spray and Wait}).

\emph{PROPHET} takes into account the frequency of past interactions
to decide on replication. However, having a high number of past interactions
does not mean nodes will have enough time and suitable conditions
to exchange information when they meet again. For instance, a stationary
node may be in the the path of a bus which results in a high frequency
of past contacts; still, speed and physical obstacles such as office
walls may lead to a problematic link seen by the proposal as a good
exchange opportunity (i.e., high delivery predictability). Thus, unwanted
copies (up to \textasciitilde{}1260 per delivery) are generated to
reach just a little over 55\% delivery probability and that take up
to 96\% (compared to \emph{Spray and Wait}) more time to reach destinations.

\emph{Epidemic}, considered as an upper bound to delivery probability,
shows a different behavior: delivery probability up to \textasciitilde{}38\%
with the highest cost per delivered message (up to \textasciitilde{}2225
copies) but with lower latency (up to \textasciitilde{}22\% compared
to \emph{Spray and Wait}). This is due to the fact that the proposal
does not worry about resource constraints, and this scenario is also
challenging given the small amount of buffer space nodes are willing
to share (i.e., 2MB).

As for the second scenario, trace-based, besides the buffer limitations,
it is even more challenging since contacts are sporadic and much less
(32) when compared to the heterogeneous scenario (962). This consequently
increases the time proposals need to have a more suitable view of
the network (in terms of the forwarding metrics they consider), thus
increasing the time to decide on replication and to deliver messages.

The advantage of \emph{Spray and Wait} reduced in this scenario where
it achieves up to 35\% of delivery with an increased cost of 15.37
copies per message delivered and latency increases going up to 44.372
seconds.

All the social-aware proposals keep a very close behavior to one another:
reaching up to 33\% (\emph{dLifeComm}) and 34\% (\emph{dLife} and
\emph{Bubble Rap}) delivery predictability with a cost per delivery
going up to \textasciitilde{}31 and 25 copies and latency increases
up to \textasciitilde{}20\% (\emph{dLifeComm}) and \textasciitilde{}24\%
(\emph{dLife} and \emph{Bubble Rap}) when compared to \emph{Spray
and Wait}.

\emph{Epidemic} and \emph{PROPHET} suffer even more with the sporadicity
of contacts in this scenario, reaching up to 19\% and 32\% for delivery
probabilities with an associated cost up to 987 and 536 copies per
successful delivery and latency increases up to \textasciitilde{}16\%
and \textasciitilde{}31\% when compared to \emph{Spray and Wait},,
respectively.

With these results one can conclude that still there is much work
to be done in the sense of creating a new opportunistic routing solution
be it based on social- oblivious or social-aware approaches. What
is more, social-aware solutions show a great potential in improving
forwarding in opportunistic networks. It is clear that there is a
tradeoff between delivery probability/cost and latency, since these
proposals take more time to choose the best next hop. It is important
to note that the performance of social-oblivious solutions such as
\emph{Spray and Wait} still require further investigation as it has
strong requirements. For instance, the number of spraying copies $L$,
during our experiments was set statically. However, according to its
paper \cite{spraywait}, this parameter depends on the number of nodes
in the network and must be estimated, and the current implementation
used in our experiments does not take into account the effect of this.

Additionally, none of the proposals consider (multi)point-to-multipoint
communication, which is a feature that should be looked upon given
the need to efficiently spread content in opportunistic networks.

As mentioned before, our goal is not to elect the best solution, but
instead to show the pros and cons of each of them, and most importantly
to show that social-aware solutions deserve attention when it comes
to developing routing approaches for opportunistic networks.

\section{Conclusions\label{sec:Conclusions}}

Social similarity has gained attention in the last years in the context
of opportunistic networks given its potential to improve data forwarding.
The reason behind this is that devices are carried by humans who happen
to have different aspects who may help to identify one or many of
them. Such aspects can be related to social ties, work affiliation,
shared interests, and among others, that can be used to infer/find
a social similarity. In addition, routing based on social similarity
has proven to be much more stable that those based on mobility update
(it is less volatile).

Thus, this chapter aims at introducing the new trend observed since
2007 with the appearance of social-aware opportunistic routing. For
that, we cover a 12-year period worth of opportunistic routing solutions
starting with a close look in the social-oblivious one to help us
in understanding the need for this new trend. Then, we follow to a
close look at pioneer social-aware solutions along with new ones,
covering how they use social similarity to devise their forwarding
approaches.

The chapter also provides a brief look at the existing opportunistic
routing taxonomies, in order to show when the social trend appeared
and how much importance is given to it. We also take this opportunity
to propose an update to an existing taxonomy, which includes the appearance
of a new sub-category of social similarity solutions based on user
dynamic behavior.

Finally, the chapter includes a set of experiments in different scenarios
pointing out advantages and disadvantages of each social-oblivious
and social-aware proposal, and indeed showing the potential of social-aware
solutions to support our case: show that the trend is strong and deserves
careful attention from researchers.

\bibliographystyle{spmpsci}
\bibliography{bib-or}

\end{document}